\documentclass[fleqn,usenatbib]{mnras}

\usepackage{mathrsfs}
\usepackage{graphicx}
\usepackage{subfigure}
\usepackage{epstopdf}
\usepackage{amsmath}
\usepackage{amssymb}
\usepackage{hyperref}
\usepackage{color}
\usepackage{bm}
\usepackage{pdflscape}

\DeclareRobustCommand{\VAN}[3]{#2}
\let\VANthebibliography\thebibliography
\def\thebibliography{\DeclareRobustCommand{\VAN}[3]{##3}\VANthebibliography}

\title[Constraints on the abundance of PBHs with different mass distributions from lensing of FRBs]{Constraints on the abundance of primordial black holes with different mass distributions from lensing of fast radio bursts}

\author[Huan Zhou et al. 2021]{
Huan Zhou,$^{1}$ Zhengxiang Li,$^{2}$\thanks{E-mail: zxli918@bnu.edu.cn}
Zhiqi Huang,$^{1}$\thanks{E-mail: huangzhq25@mail.sysu.edu.cn} He Gao,$^{2}$ Lu Huang$^{1}$
\\
$^{1}$School of Physics and Astronomy, Sun Yat-sen University, Zhuhai, 519082, China\\
$^{2}$Department of Astronomy, Beijing Normal University, Beijing 100875, China\\
}

\date{Accepted 2022 January 14. Received 2021 December 13; in original form 2021 September 29}

\pubyear{2022}

\begin{document}

\label{firstpage}
\pagerange{\pageref{firstpage}--\pageref{lastpage}}
\maketitle

\begin{abstract}
Primordial black holes (PBHs) has been considered to form a part of dark matter for a long time but the possibility  has been poorly constrained over a wide mass range, including the stellar mass range ($1-100~M_{\odot}$). However, due to the discovery of merger events of black hole binaries by LIGO-Virgo gravitational wave observatories, the interest for PBHs in the stellar mass window has been aroused again. Fast radio bursts (FRBs) are bright radio transients with millisecond duration and very high all-sky occurrence rate. Lensing effect of these bursts has been proposed as one of the optimal probes for constraining the abundance of PBHs in the stellar mass range. In this paper, we first investigate constraints on the abundance of PBHs from the latest $593$ FRB observations for both the monochromatic mass distribution and three other popular extended mass distributions related to different formation mechanisms of PBHs. It is found that constraints from currently public FRB observations are relatively weaker than those from existing gravitational wave detections. Furthermore, we forecast constraining power of future FRB observations on the abundance of PBHs with different mass distributions of PBHs and different redshift distributions of FRBs taken into account. Finally, We find that constraints of parameter space on extended mass distributions from $\sim10^5$ FRBs with $\overline{\Delta t}\leq1 ~\rm ms$ would be comparable with what can be constrained from gravitational wave events. It is foreseen that upcoming complementary multi-messenger observations will yield considerable constraints on the possibilities of PBHs in this intriguing mass window.
\end{abstract}


\section{Introduction}
The detection of the first gravitational wave (GW) event from binary black hole (BH) merger heralds the arrival of the GW astronomy era~\citep{Abbott2016}. Soon after the first detection, several research groups independently pointed out that these BHs might be primordial, i.e. primordial black holes (PBHs), and the inferred merger rate can be explained by the merger of PBHs \citep{Bird2016,Sasaki2016,Clesse2017}. PBHs can be formed in the early universe through several mechanisms, including gravitational collapse of primordial density perturbations~\citep{Hawking1971,Carr1974,Carr1975} arising from quantum fluctuation~\citep{Clesse2015,Pi2018,Ashoorioon2019,Fu2019,Cai2019,Motohashi2020}, bubble collisions~\citep{Hawking1982}, cosmic string~\citep{Hawking1989,Hogan1984}, and domain wall~\citep{Caldwell1996}. Diversity of formation mechanism means that PBHs might have different evolutionary history with respect to astrophysical black holes that are produced from the demise of massive stars. In principle, the mass of PBHs can range from the Planck mass ($10^{-5}~\rm g$) to the level of the black hole in the center of the galaxy.

Moreover, PBHs have been a field of great astrophysical interest because they are often considered to make up a part of dark matter which accounts for about one fourth of the total energy density of the universe. The scenario that the universe mainly consists of dark matter and dark energy ($\sim70\%$) is well consistent with many cosmological observations. However, we still know little about the constituent of dark matter, especially in small scales. In the past decades, intensive and continuous efforts have been conducted to search for the possibility of PBHs as dark matter candidates. Methodologically, the properties of PBHs are usually investigated via constraining the fraction of them in dark matter $f_{\rm PBH}=\Omega_{\rm PBH}/\Omega_{\rm DM}$ at present universe in various mass windows~\citep{Sasaki2018, Green2020}. These constraints could be roughly classified into two categories, i.e. direct observational constraints and indirect ones. In the direct domain, gravitational lensing~\citep{Niikura2019,Griest2013,Niikura2019a,Tisserand2007,Allsman2001,Zumalacarregui2018,Mediavilla2017,Zhou2021a}, dynamical effects on ultrafaint dwarf galaxies~\citep{Brandt2016,Koushiappas2017}, non-detections of stochastic GW~\citep{Wang2018,Clesse2017,Chen2020,Luca2020, Gert2020},  disruption of white dwarfs~\citep{Graham2015} which was shown to have huge uncertainties~\citep{Montero2019}, the accretion effects on cosmic microwave background~\citep{Haimoud2017,Aloni2017,Chen2016,Poulin2017,Bernal2017}, interstellar gas heating from observation of Leo T dwarf galaxy~\citep{Laha2021,Kim2021,Takhistov2021}, and (extra)galactic $\gamma$-ray backgrounds~\citep{Carr2010,Carr2016,Laha2019, DeRocco2019,Laha2020a,Dasgupta2020}, have been used to probe PBHs in mass windows from $\sim10^{15}~\rm g$ to supermassive scale. In the indirect domain, null detection of scalar-induced GW~\citep{Chen2019b} and cosmic microwave background (CMB) spectral distortions from the primordial density perturbations~\citep{Carr1993,Carr1994} have been proposed to derive limits on the abundance of PBHs in certain mass ranges. Recently, some other constraints from future multi-messenger observations, such as gamma-ray bursts~\citep{Ji2018}, 21 cm signals~\citep{Hektor2018,Clark2018,Halder2020}, and gravitational lensing of GW~\citep{Liao2020a,Diego2020,Urrutia2021}, have been proposed. 

PBHs within the mass range $1-100~M_{\odot}$ (stellar mass range) have attracted growing attention because of the recent detection of GW from binary black hole merger~\citep{Abbott2016} and in result GWs--a brand new observable--provide a powerful and useful tool to probe parameters of PBHs (mass, abundance etc). Meanwhile, in the electromagnetic wave domain, stellar mass PBHs also could be detected with the gravitational lensing effect of prolific transients with millisecond duration, e.g. fast radio bursts (FRBs).

FRBs are bright radio transients with short duration of few milliseconds. This mysterious burst was first reported by~\citet{Lorimer2007} and so far several hundred public FRBs are available\footnote{https://www.wis-tns.org}. Most FRBs are apparently one-off, meanwhile there are also dozens of repeaters. The observed extremely excess dispersion measure (DM: proportional to the number density of free electron along the line of sight of radio emission) of FRBs discovered at the early stage indicates their cosmological origin~\citep{Lorimer2007,Thornton2013} and it has been soon verified by the localizing the first repeater FRB121102 to a nearby dwarf galaxy~\citep{Tendulkar2017,Chatterjee2017,Marcote17}. A high event rate of this kind of mysterious phenomenon ($\sim10^3$ to $10^5$ per day all sky) has been inferred on the basis of the detection rate and the field of view of radio telescopes~\citep{Thornton2013,Champion2016,Niu2021}. For instance, it was estimated that Canadian Hydrogen Intensity Mapping Experiments (CHIME) will detect $\sim 10^4$ FRBs per year, of which $\sim10^3$ FRBs with redshift information will be probed~\citep{Connor2016}. 
There are still intensive debates in radiation mechanism and progenitors of these mysterious bursts. Recent progress in detection of a Galactic FRB in association with a soft gamma-ray repeater suggests that magnetar engines can produce at least some (or probably all) FRBs~\citep{Zhang2020,CHIME/FRB2020,Bochenek2020,Lin2020}. However, some unique and useful observational properties of FRBs, including clean temporal shape, short temporal duration, cosmological origin, and high all-sky event rate have been proposed as promising cosmological and astrophysical probes, such as testing fundamental physics~\citep{Wei2015,Wu2016}, constraining cosmological models ~\citep{Gao2014,Zhou2014,Walters2018,Zhao2020}, baryon census~\citep{Deng2014,Munoz2018,Macquart2020,Li2019,Li2020}, reionization history of universe~\citep{Linder2020,Bhattacharya2020,Beniamini2021}, millilensed lensed FRBs for probing compact dark matter ~\citep{Munoz2016,Wang2018a,Liao2020,Laha2020,Katz2020,Zhou2021}, galaxy lensing time delay variations for probing the the motion of the FRB source~\citep{Dai2017}, and time delay distances of strongly lensed FRBs for precisely measuring the expansion rate and curvature of the universe~\citep{Li2018,Wucknitz2020}.

In this paper, following the method proposed in~\citet{Liao2020,Laha2020}, we concentrate on constraining the abundance of PBHs in the stellar mass range from lensing of FRBs. First, with the monochromatic mass distribution (MMD) and three other extended mass distributions (EMDs), we derive constraints on the the abundance of PBHs from the null search result of the latest FRB observations. Moreover, we investigate constraints on the PBH properties from near future FRB observations by taking the detection ability of already running and upcoming telescopes and possible redshift distribution of FRBs into consideration. Finally, comparisons between constraints on the PBH properties from FRBs observations and the ones obtained from GW detection are presented to explore the possibility of deriving joined constraints on PBHs from upcoming multi-messenger observations.

This paper is organized as follows. We review the theory of FRBs lensing by PBHs with different mass functions in Sec.~\ref{sec2}. In Sec.~\ref{sec3} we briefly introduce the latest FRB observations and present the constraints on PBHs from them for both MMD and EMDs. Sec.~\ref{sec4} shows constraints from upcoming FRB observations on parameter space of PBHs with different EMD, as well as comparisons between the constraints on PBHs from the merger rate of PBH binaries in view of GW detection with the ones from near future FRB observations. Finally, we conclude and discuss in Sec.~\ref{sec5}.

\section{Lensing of FRBs}\label{sec2}
The theory of gravitational lensing is a well-studied phenomenon and is used as an observational tool in astrophysics to probe objects which are too faint to be detected by conventional means. It has been proposed for probing PBHs in the universe for a long time. For a lensing system, Einstein radius is one of the characteristic parameters and, taking the intervening lens as a point mass, it is given by
\begin{equation}\label{eq1}
\theta_{\rm E}=2\sqrt{\frac{GM_{\rm PBH}D_{\rm LS}}{c^2D_{\rm L}D_{\rm S}}}\approx (3\times 10^{-6})^{''}\bigg(\frac{M_{\rm PBH}}{M_{\odot}}\bigg)^{\frac{1}{2}}\bigg(\frac{D}{\rm Gpc}\bigg)^{-\frac{1}{2}},
\end{equation}
where $G$ and $c$ represent the gravitational constant and the speed of light, respectively. In addition, $D=D_{\rm L}D_{\rm S}/D_{\rm LS}$ is effective lensing distance, where $D_{\rm S}$, $D_{\rm L}$ and $D_{\rm LS}$ represent the angular diameter distance to the source, to the lens, and between the source and the lens, respectively. Although the angular resolution for some repeaters with very long baseline array (VLBA) could reach a high level, e.g. $\sim(10^{-2})^{''}$ of VLBA observations for FRB 121102  \citep{Spitler2016,Chatterjee2017,Tendulkar2017,Marcote17}, it is just possible to distinguish multiple images of an FRB lensed by intervening objects with mass greater than $10^8M_{\odot}$. 
Another characteristic parameter for a lensing system is the time delay between lensed signals and it can be approximately expressed as,
\begin{equation}\label{eq2}
\Delta t\approx 1~{\rm ms}~\bigg(\frac{M_{\rm PBH}}{30~M_{\odot}}\bigg).
\end{equation}
Therefore, for lensing systems where PBHs with mass in $1-100~M_{\odot}$ acting as deflectors, the time delay between multiple images is $\sim 0.1-1~\rm ms$ and is accurately determined by,
\begin{equation}\label{eq3}
\begin{split}
\Delta t(M_{\rm PBH},z_{\rm L},y)=\frac{4GM_{\rm PBH}}{c^3}\big(1+z_{\rm L}\big)\\
\bigg[\frac{y}{2}\sqrt{y^2+4}+\ln\bigg(\frac{\sqrt{y^2+4}+y}{\sqrt{y^2+4}-y}\bigg)\bigg],
\end{split}
\end{equation}
where the normalized impact parameter $y=\beta/\theta_{\rm E}$ is defined as the ratio of the angular impact parameter to the angular Einstein radius, $z_{\rm L}$ is the lens redshift. $\Delta t$ must be larger than the width ($w$) of the observed signal to resolve lensed echoes. This requires $y$ larger than a certain value $y_{\min}(M_{\rm PBH},z_{\rm L},w)$ according to Eq.~\ref{eq3}. The lensing cross section due to a PBH lens is given by an annulus between the maximum and minimum impact parameters,
\begin{equation}\label{eq4}
\begin{split}
\sigma(M_{\rm PBH}, z_{\rm L}, z_{\rm S}, w)=\frac{4\pi GM_{\rm PBH}D_{\rm L}D_{\rm LS}}{c^2D_{\rm S}}\\
[y^2_{\max}(R_{\rm f})-y^2_{\min}(M_{\rm PBH},z_{\rm L},w)],
\end{split}
\end{equation}
where $R_{\rm f}$ is the flux ratio. It is defined as the absolute value of the ratio of the magnifications $\mu_+$ and $\mu_-$ of both images,
\begin{equation}\label{eq5}
R_{\rm f}\equiv\bigg|\frac{\mu_+}{\mu_-}\bigg|=\frac{y^2+2+y\sqrt{y^2+4}}{y^2+2-y\sqrt{y^2+4}}>1.
\end{equation}
The maximum value of normalized impact parameter can be found by requiring that the flux ratio of two lensed images is smaller than a critical value $R_{\rm f, max}$,
\begin{equation}\label{eq6}
y_{\max}(R_{\rm f,max})=R_{\rm f,max}^{1/4}-R_{\rm f,max}^{-1/4},
\end{equation}
to ensure that both signals are detectable. Here, following~\citet{Munoz2016}, we take $R_{\rm f,max}=5$ for cases when we study lensing of the whole sample of all currently public FRBs. For a single source, the optical depth for lensing due to a single PBH is
\begin{equation}\label{eq7}
\begin{split}
\tau(M_{\rm PBH},f_{\rm PBH},z_{\rm S},w)=\int_0^{z_{\rm S}}d\chi(z_{\rm L})(1+z_{\rm L})^2n_{\rm L}(f_{\rm PBH})\\
\sigma(M_{\rm PBH},z_{\rm L},z_{\rm S}, w)=
\frac{3}{2}f_{\rm PBH}\Omega_{\rm DM}\int_0^{z_{\rm S}}dz_{\rm L}\frac{H_0^2}{cH(z_{\rm L})}\\
\frac{D_{\rm L}D_{\rm LS}}{D_{\rm S}}(1+z_{\rm L})^2[y^2_{\max}(R_{\rm f,max})-y^2_{\min}(M_{\rm PBH},z_{\rm L},w)],
\end{split}
\end{equation}
where $n_{\rm L}$ is the comoving number density of the lens, $H(z_{\rm L})$ is the Hubble parameter at $z_{\rm L}$, $H_0$ is the Hubble constant and we use the value estimated from the latest CMB observations~\citep{Planck2018} in our following analysis, $f_{\rm PBH}$ represents the fraction of PBHs in dark matter, and $\Omega_{\rm DM}$ is the present density parameter of dark matter. In order to find the total lensing optical depth, one has to integrate the optical depth in Eq.~\ref{eq7}. At present, the latest observations of FRB consist of $593$ well-identified events, which provides a statistically meaningful sample. According to the definition, the expected number of lensed FRBs in currently public observations is approximately equivalent to the sum of the lensing optical depths of all FRBs,
\begin{equation}\label{eq8}
N_{\rm lensed~FRB}=\sum_{i=1}^{N_{\rm total}}\tau_i(M_{\rm PBH},f_{\rm PBH},z_{S,i},w_i).
\end{equation}

It should be pointed out that the above formalism is only valid for the simple but often used MMD,
\begin{equation}\label{eq9}
P(m,M_{\rm PBH})=\delta(m-M_{\rm PBH}),
\end{equation}
where $\delta(m-M_{\rm PBH})$ represents the $\delta$-function at the mass $M_{\rm PBH}$. In fact, there are some EMDs which have more robust physical motivations. Therefore, it is important and necessary to derive constraints on PBH with some theoretically motivated EMDs, which are closely related to formation mechanisms of PBHs. In this paper, we take three EMDs into consideration, i.e. the extend power-law mass function, the log-normal mass function, and the critical collapse mass function. Concretely, the extend power-law mass function is parametrized as~\citep{Laha2020,Bellomo2018},
\begin{equation}\label{eq10}
P(m,M_{\min},M_{\max},\gamma)=\frac{\mathscr{N}_{\rm pl}}{M^{1-\gamma}}\Theta(M_{\max}-m)\Theta(m-M_{\min}),
\end{equation}
where the mass range of the distribution is bordered by the minimum mass, $M_{\min}$, and the maximum mass, $M_{\max}$. The exponent of the power law is denoted by $\gamma$. The normalization factor $\mathscr{N}_{\rm pl}$ is,
\begin{equation}\label{eq11}
\mathscr{N}_{\rm pl}=\left\{
\begin{aligned}
\frac{\gamma}{M_{\max}^{\gamma}-M_{\min}^{\gamma}},\quad\gamma\neq0,\\
\frac{1}{\ln(M_{\max}/M_{\min})},\quad\gamma=0,
\end{aligned}
\right.
\end{equation}
where the exponent $\gamma$ is determined by the formation epoch of the PBH, and the natural range of exponent $\gamma$ is $\gamma\in(-1,1)$, which corresponds to the equation of state $\omega\in(-1/3,1)$. Interesting values of the exponent are $\gamma=-0.5$ and $\gamma=0$, corresponding to formation during radiation and matter dominated, respectively. We take $\gamma$ as a free parameter with prior $\gamma\leq0$. This mass function can arise if the PBHs are generated by scale-invariant density fluctuations or from the collapse of cosmic strings. The log-normal mass function is~\citep{Gert2020, Carr2017, Bellomo2018}:
\begin{equation}\label{eq12}
P(m,\sigma,m_{\rm c})=\frac{1}{\sqrt{2\pi}\sigma m}\exp\bigg(-\frac{\ln^2(m/m_{\rm c})}{2\sigma^2}\bigg),
\end{equation}
where $m_{\rm c}$ and $\sigma$ denote the peak mass of $mP(m)$ and the width of mass spectrum, respectively. This mass function is often a good approximation if the PBHs produced from a smooth symmetric peak in the inflationary power spectrum, and it had been demonstrated numerically in~\cite{Green2016} and analytically in~\cite{Kannike2017} for the case where the slow-roll approximation holds. Therefore, it is a representative of a large class of extend mass functions. A critical collapse mass function is~\citep{Carr2017, Gert2020, Carr2016}:
\begin{equation}\label{eq13}
P(m,\alpha,M_{\rm f})=\frac{\alpha^2}{M_{\rm f}^{1+\alpha}\Gamma(1/\alpha)}m^{\alpha}\exp(-(\frac{m}{M_{\rm f}})^{\alpha}),
\end{equation}
where $M_{\rm f}$ is a mass-scale which corresponds to the horizon mass at the collapse epoch, and $\alpha\sim3$ is a universal exponent which is related to the critical collapse of radiation. In this case, the mass spectrum can extend down to arbitrarily low masses, but there is an exponential upper cut-off at a mass-scale $M_{\rm f}$. We take $\alpha$ as a free parameter ranging from 1.5 to 3 in our analysis. This mass function is supposed to be closely related to PBHs originating from density fluctuations with a $\delta$-function power spectrum.

For the above-mentioned EMDs, the lensing optical depth for a given FRB can be written as,
\begin{equation}\label{eq14}
\begin{split}
\tau(f_{\rm PBH},z_{\rm S},w, \boldsymbol p_{\rm mf})=\int dm\int_0^{z_{\rm S}}d\chi(z_{\rm L})(1+z_{\rm L})^2\\
n_{\rm L}(f_{\rm PBH})\sigma(m,z_{\rm L},z_{\rm S}, w)P(m, \boldsymbol p_{\rm mf}),
\end{split}
\end{equation}
where $\boldsymbol p_{\rm mf}$ represents the set of parameters in the mass function. Obviously,  the optical depth for the MMD of PBHs  Eq.~\ref{eq7} can be derived by combining Eq.~\ref{eq14} and Eq.~\ref{eq9}.

Intuitively, these constraints can be understood by following the equivalent mass formalism introduced in\citep{Bellomo2018, Laha2020}. In this formalism, one finds a single lens mass which represents the full effect of the EMD. This formalism is valid for any EMD and we can use it to recast the limits on the PBHs that had been derived for MMD. We can calculate the equivalent mass for the EMDs by equating the expressions for the integrated optical depth with appropriate changes:
\begin{equation}\label{eq15}
\begin{split}
\int dz_{\rm sm}\tau(M_{\rm PBH}^{\rm eq}, f_{\rm PBH}^{\rm MMD}, z_{\rm sm}, w)N(z_{\rm sm})\\
=\int dz_{\rm se}\tau(f_{\rm PBH}^{\rm EMD}, z_{\rm se}, w, \boldsymbol p_{\rm mf})N(z_{\rm se}),
\end{split}
\end{equation}
where the equivalent mass is denoted by $M^{\rm eq}_{\rm PBH}$ when $f_{\rm PBH}^{\rm MMD}=f_{\rm PBH}^{\rm EMD}$ ($f_{\rm PBH}^{\rm MMD}$ and $f_{\rm PBH}^{\rm EMD}$ represent the fraction of PBHs when MMD and EMD are used respectively). The redshifts for the MMD and EMD are denoted by $z_{\rm sm}$ and $z_{\rm se}$, respectively. After deducing the equivalent mass for a given input of EMD, we can then read off the $f_{\rm PBH}$ for that distribution by using Figs. (\ref{fig3},\ref{fig4}).

\section{constraints from the latest FRB observations}\label{sec3}
In this section, we first describe the latest FRB data. In addition, we show the constraints on the abundance of PBHs from the latest FRBs observations with both MMD and EMDs taken into account.

\subsection{The latest FRB data}\label{sec3.1}
\begin{figure}
    \centering
     \includegraphics[width=0.45\textwidth, height=0.3\textwidth]{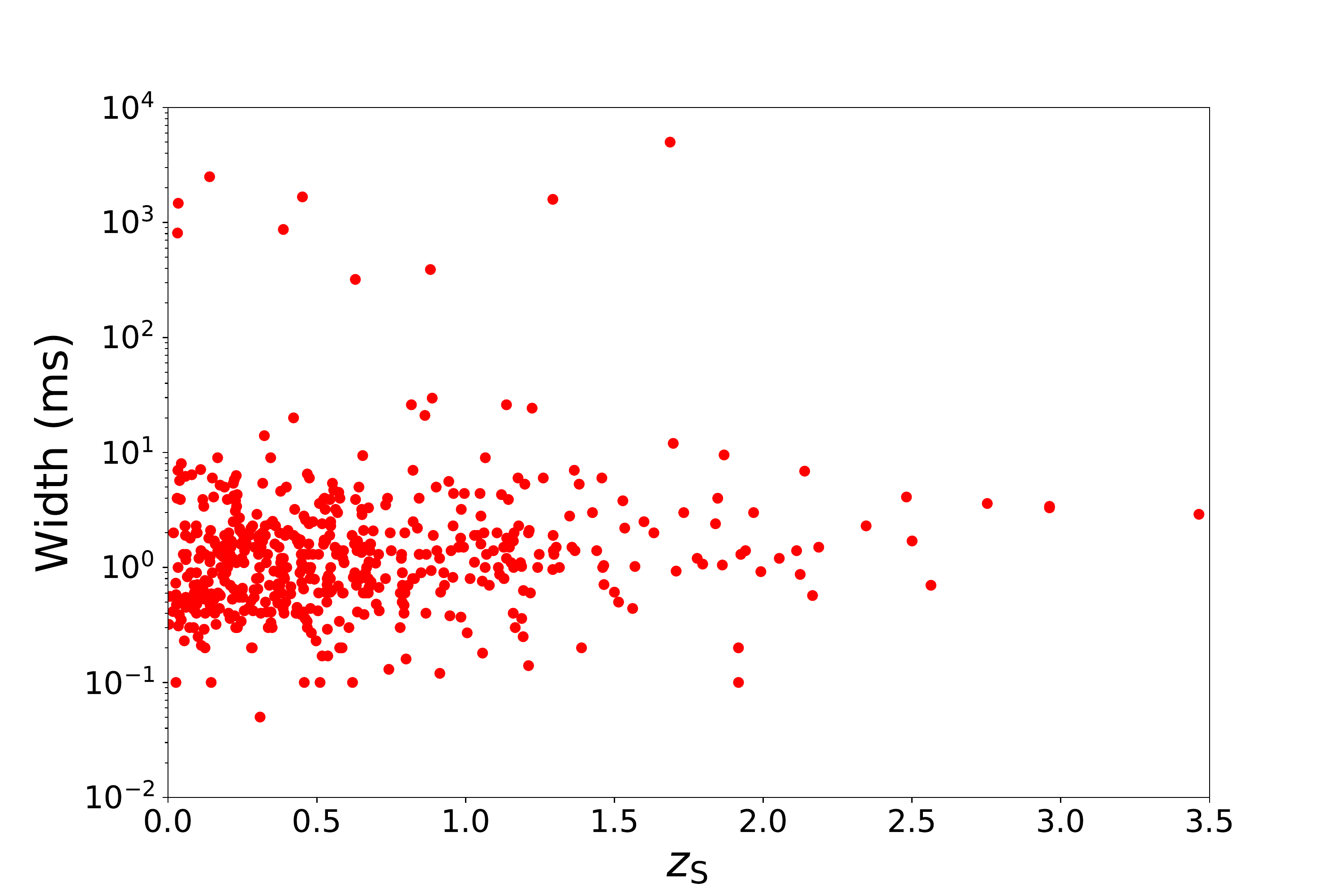}
     \caption{Two-dimensional distribution of inferred redshifts and widths for the latest 593 FRBs.}\label{fig1}
\end{figure}

At present, there are about $593$ publicly available FRBs. This number has significantly increased because of the release of the first CHIME/FRB FRB catalog, which consist of more than five hundred events detected in less than one year (2018 July 25 to 2019 July 1)\footnote{https://www.chime-frb.ca/catalog}~\citep{CHIME/FRB2021}. For a detected FRB, one of the most important observational features is the DM, which is measured from the delayed arrival time of two photons with different frequencies. Even though the first several bursts were poorly localized~\citep{Lorimer2007,Thornton2013}, their cosmological origin was inferred from their extremely excess observed DMs compared with Milky Way DM contributions in the same directions. As expected, their extragalactic origin was soon confirmed by the localization of the first repeater FRB 121102 to a dwarf galaxy at $z\sim0.19$~\citep{Tendulkar2017,Chatterjee2017,Marcote17}. As a result, the distance and redshift of a detected FRB can be approximately estimated from its observed DM which is usually decomposed into the following four ingredients,
\begin{equation}\label{eq16}
{\rm DM}=\frac{\rm DM_{\rm host}+DM_{\rm src}}{1+z}+{\rm DM_{IGM}}+{\rm DM_{MW}},
\end{equation}
where ${\rm DM_{host}}$ and ${\rm DM_{src}}$ represent DM from host galaxy and local environment, respectively. These two components are relatively uncertain and significantly affect the redshift estimation from DM measurements. That is, assuming a large host contribution leads to a low inferred redshift. Here, we adopt the minimum inference of redshift for all host galaxies, which corresponds to the maximum value of ${\rm DM_{host}}+{\rm DM_{src}}$ to be 200 $\rm pc/cm^{3}$. Moreover, we also assume this term being $50~\rm pc/cm^{3}$ and $100~\rm pc/cm^{3}$ to investigate the influence of the uncertainty of these two components on our following analysis. ${\rm DM_{MW}}$ is the contribution from the Milky Way. In addition, the intergalactic medium (IGM) contribution ${\rm DM_{IGM}}$ is closely related to the redshift or distance of the source and the baryon content of the universe. The ${\rm DM_{IGM}}-z$ relation is given by~\citep{Deng2014} and it is approximately expressed as ${\rm DM_{IGM}}\sim855z~\rm pc/cm^3$ by considering the fraction $f_{\rm IGM}$ of baryon in the IGM to $f_{\rm IGM}=0.83$ and the He ionization history~\citep{Zhang2018}. DM and redshift measurements for several localized FRBs suggested that this relation is statistically favored by observations~\citep{Li2020}. Duration of FRB is also an important observational feature for determining the mass range of PBHs to be constrained from the lensing effect of these bursts, so we collect the inferred redshifts and pulse widths of all currently available FRBs and present them in Fig.~\ref{fig1}. 

Moreover, for the latest $593$ available FRBs, especially those with multiple peaks, we have carefully checked their light curves, intrinsic structures, and dynamic spectra, and no candidate with strong evidence of lensing effect has been found~\citep{Zhou2021}. In our following analysis, we derive constraints on the abundance of PBHs from this null search result of lensed signals with different mass distributions taken into consideration.

\subsection{Constraints from the latest data}\label{sec3.2}
\begin{figure*}
\centering
\subfigure[Monochromatic mass function]{\includegraphics[width=0.45\textwidth, height=0.3\textwidth]{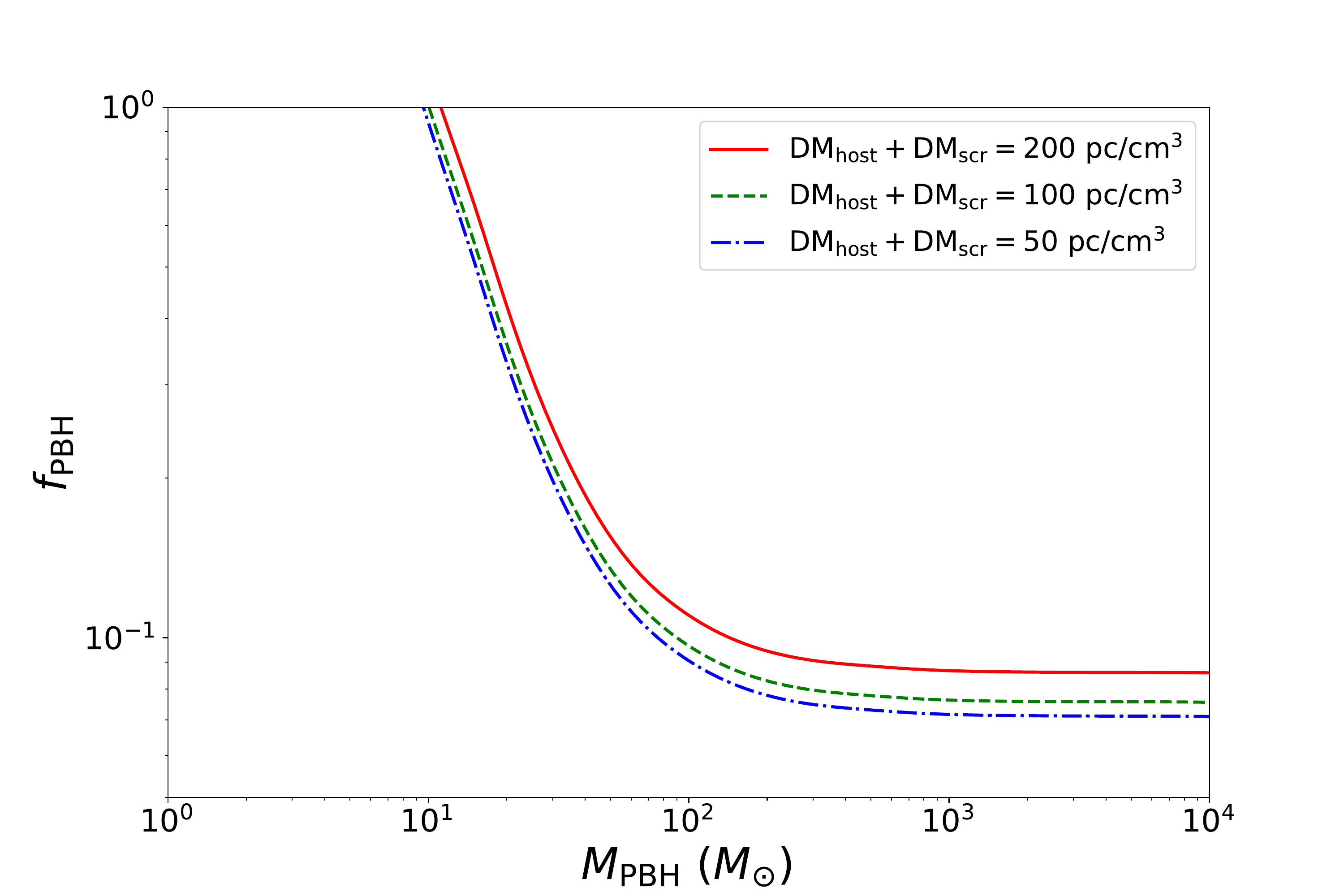}}
\subfigure[Extend power-law mass function ]{\includegraphics[width=0.45\textwidth, height=0.3\textwidth]{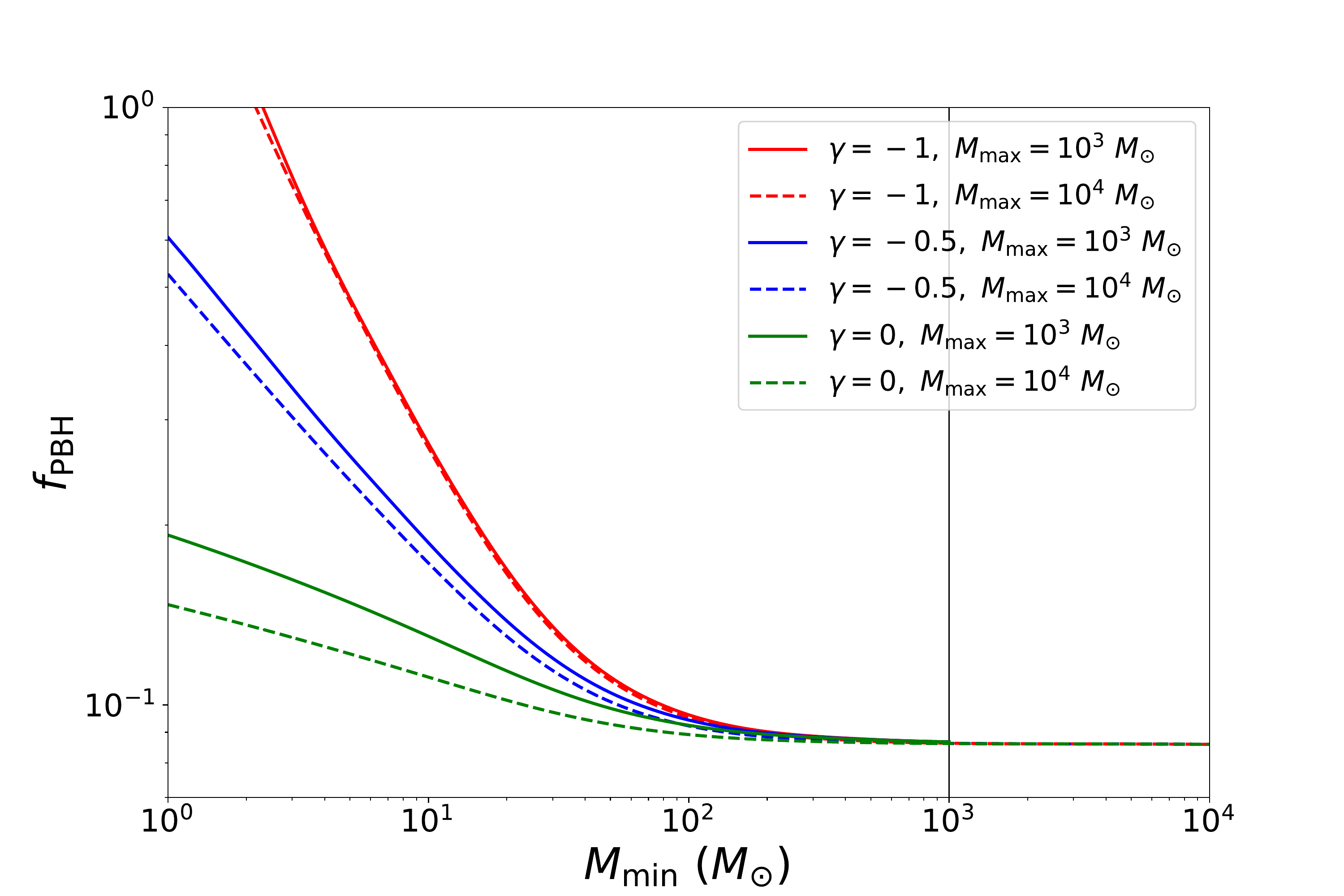}}
\subfigure[Log-normal mass function]{\includegraphics[width=0.45\textwidth, height=0.3\textwidth]{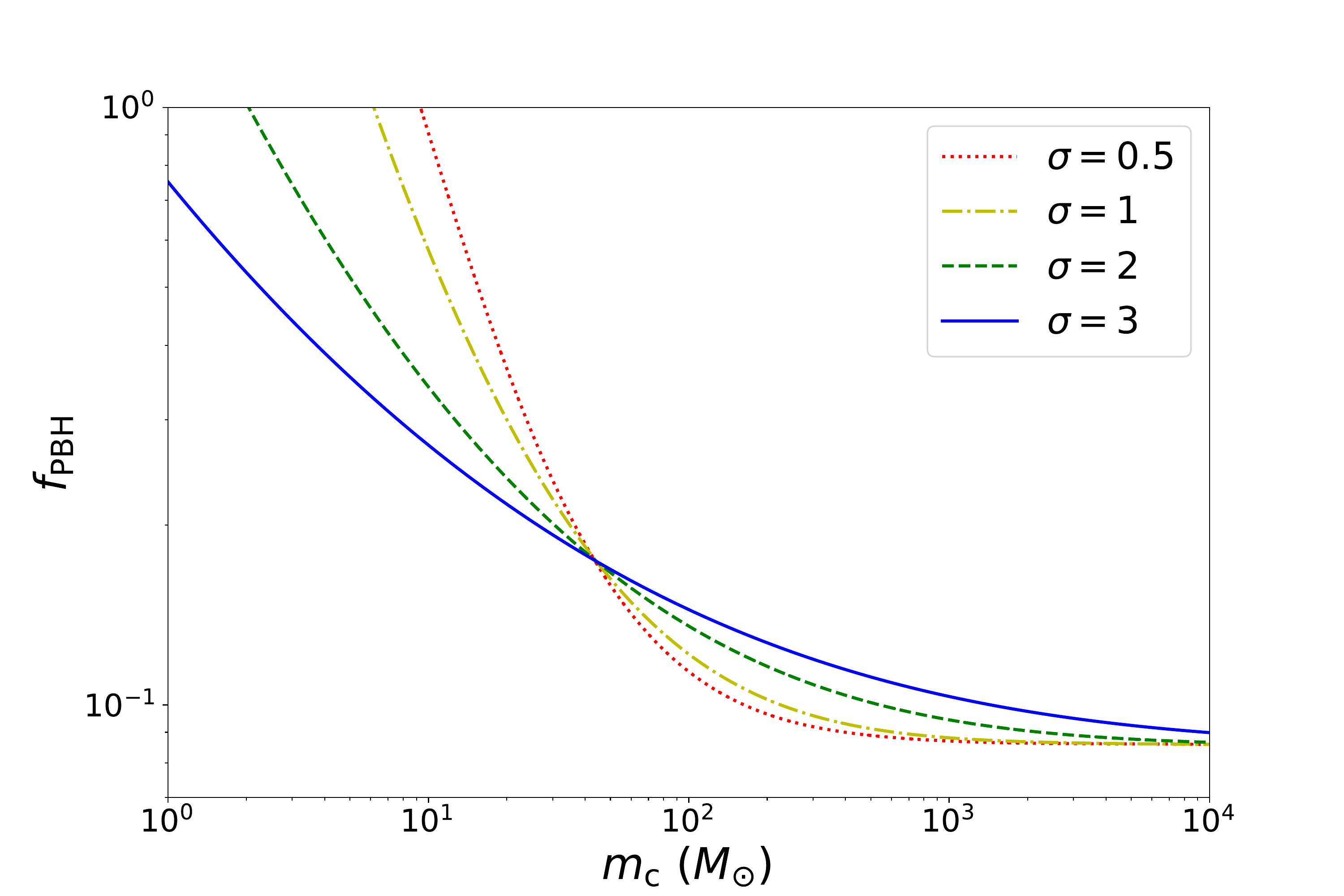}}
\subfigure[Critical collapse mass function]{\includegraphics[width=0.45\textwidth, height=0.3\textwidth]{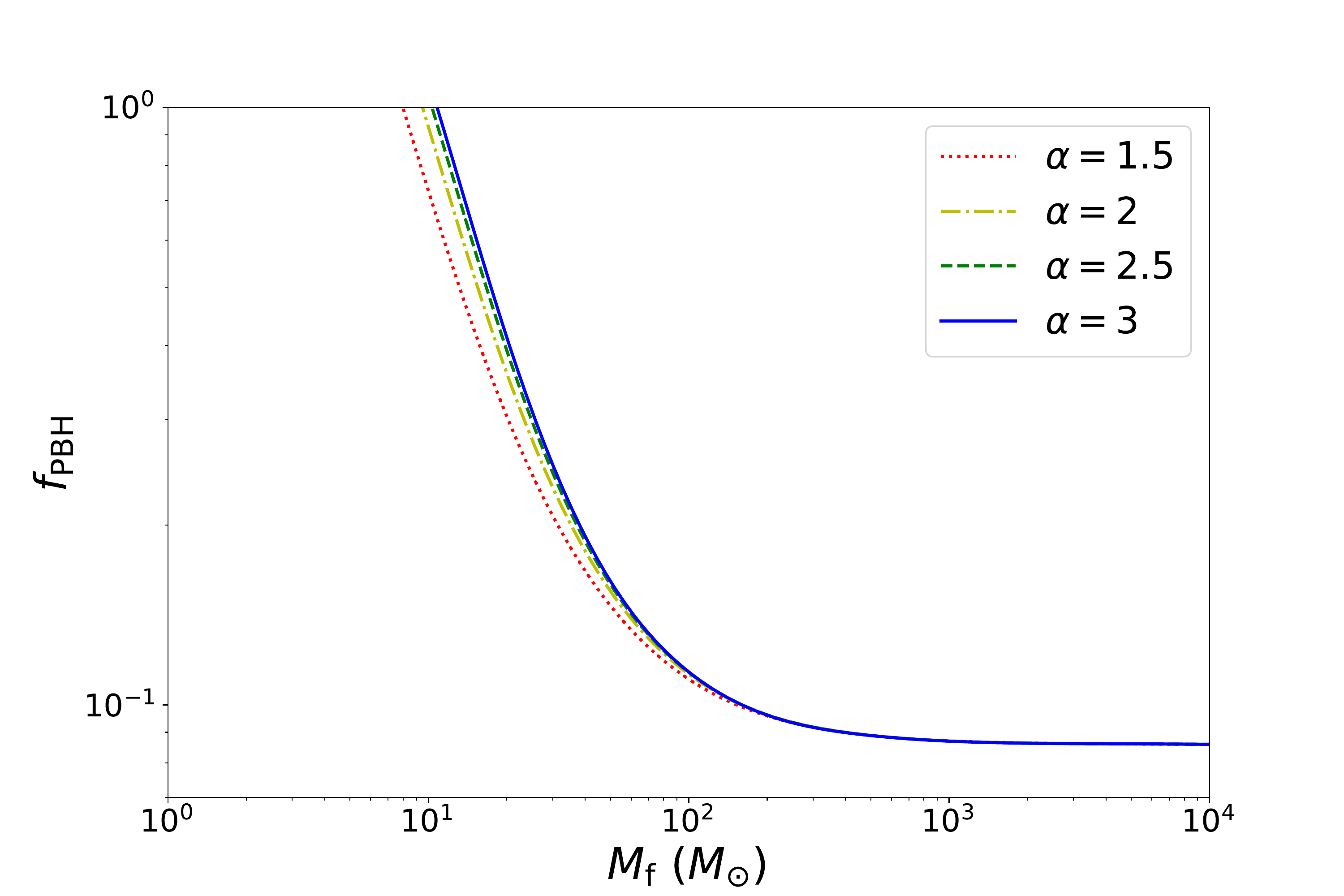}}
\caption{{\bf Upper left:} Constraints on the upper limits of fraction of dark matter in the form of PBHs with the MMD function from the fact that no lensed signal has been found in $593$ FRBs data. {\bf Upper right:} Same as the upper left panel but for the extend power-law distribution with $\gamma=0$, $-0.5$, and $-1$ and a fixed $M_{\max}$. Solid lines represent the constraints of $M_{\max}=10^3~M_{\odot}$ and dashed lines represent the constraints of $M_{\max}=10^4~M_{\odot}$. {\bf Lower left:} Same as the upper left panel but for the log-normal distribution with $\sigma=0.5$, 1, 2, and 3. {\bf Lower right:} Same as the upper left panel but for the critical collapse distribution for $\alpha=1.5$, $2$, $2.5$, and $3$.}\label{fig2}
\end{figure*}

We first derive constraints on the abundance of PBHs from this null search result of lensed FRBs in the framework of MMD. Each $(M_{\rm PBH},f_{\rm PBH})$ corresponds to an expected number of lensed FRB signals according to Eqs. (\ref{eq7}, \ref{eq8}). Since no lensed FRB event has been found in the current data, the curve in the $(M_{\rm PBH},f_{\rm PBH})$ parameter space that predict one detectable lensed FRB, as shown in panel~(a) of Fig.~\ref{fig2}, is disfavored at 63\% confidence level (assuming Poisson events and uniform prior on the summed lensing optical depth). In the $\lesssim 10~M_{\odot}$ low-mass end, $f_{\rm PBH}$ is unconstrained because of the time delay ($\Delta t$) resolution limit; For the mass range $10-10^3~M_{\odot}$, the constraint gradually improves towards larger mass end, where the $\Delta t$ resolution impact diminishes. In the $\gtrsim 10^3~M_{\odot}$ large-mass end, the constraint on $f_{\rm PBH}$ saturates to $\lesssim 7.1\%$, $\lesssim 7.6\%$, and $\lesssim 8.6\%$ at 63\% confidence level when the host contribution of DM is assumed to be $50~\rm pc/cm^{3}$, $100~\rm pc/cm^{3}$, and $200~\rm pc/cm^{3}$, respectively. The overall result is improved compared to the constraint on the compact objects presented in~\citet{Liao2020}.

Next, we derive constraints on $f_{\rm PBH}$ assuming that masses of PBHs follow the extended power-law distribution, and the results is shown in panel~(b) of Fig.~\ref{fig2}. Here, we assume $M_{\max}$ in the extended power-law mass function to be $M_{\max}=10^3~M_{\odot}$ (or $10^4~M_{\odot}$) and vary the minimum value, $M_{\min}$ from $1~M_{\odot}$ to $999~M_{\odot}$ (or $9999~M_{\odot}$). In addition, we set three values for the parameter $\gamma$, i.e. $0$, $-0.5$, and $-1$. Analogously, each $(M_{\min},f_{\rm PBH})$ corresponds to an expected number of lensed FRB signals according to Eqs. (\ref{eq8}, \ref{eq14}). Again we show the curves that predict one detectable lensed signal as references. In this model, the constraints from larger mass end is extrapolated to the low-mass end by the assumed power-law form. The $\gamma=0$ case corresponds to a flatter mass distribution, which leads to a better extrapolation efficiency. Similar to the MMD model, the $M_{\max}=10^3~M_{\odot}$ cases are limited by the $\Delta t$ resolution limit, and hence $f_{\rm PBH}$ are typically worse constrained than the $M_{\max}=10^4~M_{\odot}$ cases.

For the log-normal mass distribution, the results of the projected constraints on $f_{\rm PBH}$ is shown in the lower left panel of Fig.~\ref{fig2}. We set the width of mass spectrum $\sigma$ to be several typical values: $0.5$, $1$, $2$, and $3$. In addition, we assume the value of the peak mass $m_{\rm c}$ varies from $1~M_{\odot}$ to $10^4~M_{\odot}$. Similarly, each $(m_{\rm c},f_{\rm PBH})$ corresponds to an expected number of lensed FRB signals according to Eqs. (\ref{eq8}, \ref{eq14}), and the area in the $(m_{\rm c},f_{\rm PBH})$ parameter space above the curve which predicts one detectable lensed signal should be disfavored because no lensed FRB event has been found in the current data. For a fixed $m_c\gtrsim 45~M_{\odot}$, a broader mass distribution (larger $\sigma$) typically leads a worse constraint on $f_{\rm PBH}$. This is because more $\Delta t$-unresolved (low-$M_{\rm PBH}$) lensing cases are allowed in a broader distribution. This argument, however, cannot be applied if $m_c \lesssim 45~M_{\odot}$, where a broader mass distribution also brings in more $45~M_{\odot}$ samples that are better constrained than the $M_{\rm PBH}=m_c$ case. We observe numerically that this effect is not competitive enough to beat the previous mentioned effect that extension to lower-mass end worsens the constraint.

For the critical collapse mass distribution, the constraints on $f_{\rm PBH}$ from the latest FRB observations is shown in the lower right panel of Fig.~\ref{fig2}. In this scenario, the value of $\alpha$, which is a universal exponent relating to the critical collapse of radiation, is assumed to be $1.5$, $2$, $2.5$, and $3$ and the mass-scale $M_{\rm f}$ is assumed to vary from $1~M_{\odot}$ to $10^4~M_{\odot}$. Similar to the above-mentioned three cases, each $(M_{\rm f},f_{\rm PBH})$ corresponds to an expected number of lensed FRB signals according to Eqs. (\ref{eq8}, \ref{eq14}). Since no lensed signal has been found in the current data, in the $(M_{\rm f},f_{\rm PBH})$ parameter space, the region above the curve that predicts one detectable lensed signal is disfavored. For large $\alpha$'s, this mass distributions, and hence the constraints on $f_{\rm PBH}$, are all very close to the MMD case. For a smaller $\alpha$, the mass distribution is flatter in the small mass range. Similar to the log-normal case, for a flatter mass distribution two effects compete. The outreach to the larger-mass end improves the constraint on $f_{\rm PBH}$, whereas the extension to the lower-mass end worsens the constraint. The numeric result here suggests that, in contrary to the log-normal case, here flattening of distribution slightly improves the constraint. 

\section{Forecasts}\label{sec4}

Although the current constraints from the latest observations are weak due to a small number of FRBs  available at this moment, and this work is more like a proof of concept, more stringent limits on PBH properties will be achieved from upcoming observations of this new and promising probe, owing to rapid progress in the FRB observation community.

It is expected that upcoming wide field radio surveys, especially CHIME \citep{Chime2018}, ASKAP \citep{Askap2019}, DSA-2000 \citep{Hallinan2019}, will detect a large number of FRBs, i.e. $\sim10^4$ per year. In addition, several other telescopes, like the Five-hundred-meter Aperture Spherical Telescope (FAST) \citep{Zhang2018}, the Upgraded Giant Metrewave Radio Telescope~\citep{Bhattacharyya2018}, Ooty Wide Field Array~\citep{Bhattacharyya2018}, UTMOST~\citep{Bailes2017}, HIRAX~\citep{Weltman2017}, and APERTIF~\citep{Maan2017} will revolutionize the fields of FRB observations and applications. 

If we accumulate a considerable number of FRBs with redshifts satisfying $N(z_{\rm S})$ distribution in the near future, the integrated optical depth of all these bursts $\bar{\tau}(M_{\rm PBH},f_{\rm PBH},w)$ is,
\begin{equation}\label{eq17}
\bar{\tau}(M_{\rm PBH},f_{\rm PBH},w)=\int dz_{\rm S}\tau(M_{\rm PBH},f_{\rm PBH},z_{\rm S},w)N(z_{\rm S}).
\end{equation}
Consequently, the expected number of lensed FRBs is,
\begin{equation}\label{eq18}
N_{\rm lensed~FRB}=(1-e^{-\bar{\tau}(M_{\rm PBH},f_{\rm PBH},w)})N_{\rm FRB}.
\end{equation}
In our following analysis, we use two realistic redshift distributions $N(z)$ of FRBs to make further forecasts, i.e. the constant-density redshift distribution (CRD) $N_{\rm CRD}(z)$ and the star-formation redshift distribution (SRD) $N_{\rm SRD}(z)$. Specifically, the CRD is expressed as~\citep{Munoz2016,Laha2020,Oppermann2016},
\begin{equation}\label{eq19}
N_{\rm CRD}(z)=\mathscr{N}_{\rm c}\frac{\chi^2(z)e^{-d^2_{\rm L}(z)/[2d^2_{\rm L}(z_{\rm cut})]}}{(1+z)H(z)},
\end{equation}
where $\mathscr{N}_{\rm c}$ is a normalization factor to ensure that $N_{\rm CRD}(z)$ integrates to unity, $d_{\rm L}(z)$is the luminosity distance, in this work computed with Planck best-fit $\Lambda$CDM model~\citep{Planck2018}, and $z_{\rm cut}$ is a Gaussian cutoff in the FRB redshift distribution due to an instrumental signal-to-noise threshold. The SRD can be written as~\citep{Munoz2016,Laha2020,Caleb2016}:
\begin{equation}\label{eq20}
N_{\rm SRD}=\mathscr{N}_{\rm s}\frac{\dot{\rho}_*(z)\chi^2(z)e^{-d^2_{\rm L}(z)/[2d^2_{\rm L}(z_{\rm cut})]}}{(1+z)H(z)},
\end{equation}
where $\mathscr{N}_{\rm s}$ is the normalization constant which is determined from $\int dzN_{\rm SRD}(z)=1$, and
\begin{equation}\label{eq21}
\dot{\rho}_*(z)=h\frac{a+bz}{1+(z/s)^d},
\end{equation}
where $a=0.017$, $b=0.13$, $s=3.3$, $d=5.3$, and $h=0.7$. It is shown that, compared with the CRD distribution, SRD is slightly preferred~\citep{James2021}. Besides these two commonly used distributions, \citet{Qiang2021} also studied many other redshift distributions which are consistent with data.

In this section, we firstly present forecasts of constraints on PBH properties from upcoming FRBs assuming two redshift distributions (CRD and SRD) of them. Moreover, we compare the forecasts with the constraints from GW observations to study the possibility of joint constraints on PBHs from near future multi-messenger observations.

\subsection{Forecast with different mass function}\label{sec4.1}
\begin{figure}
    \centering
     \includegraphics[width=0.45\textwidth, height=0.3\textwidth]{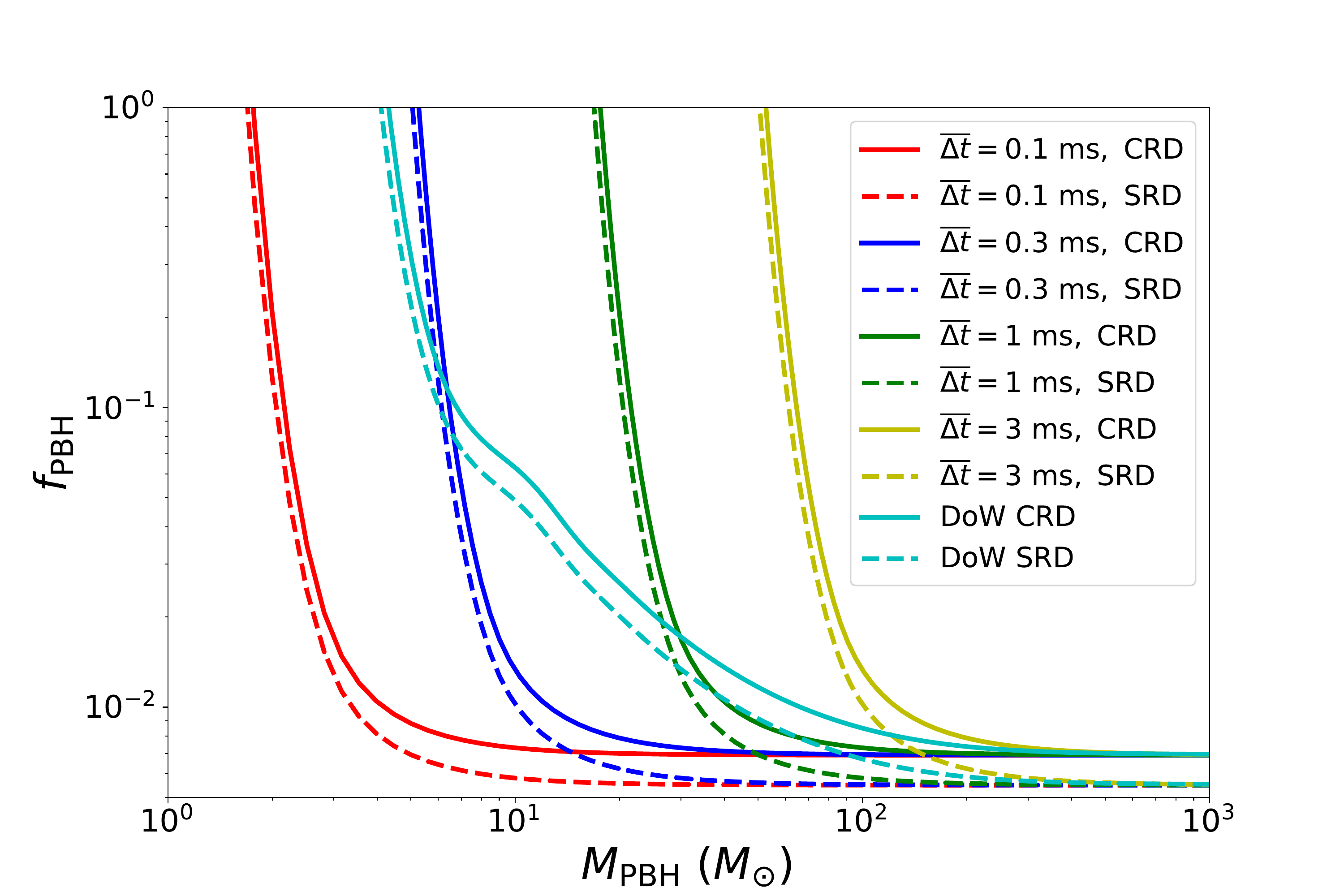}
     \caption{Constraints on the upper limits of the fraction of dark matter in the form of PBH with the MMD from the fact that no lensed signal has been found in the upcoming $10^4$ FRBs with average values of the critical time, $\overline{\Delta t}=0.1,$ 0.3, 1, and $3~\rm~ms$. DoW represents the forecast with $10^4$ no lensed FRBs based on the current distribution of widths. The solid lines and dashed lines represent results when the CRD and SRD is considered, respectively.}\label{fig3}
\end{figure}

In Fig.~\ref{fig3}, we demonstrate the constraints on dark matter fraction in the form of PBHs when the MMD is considered. These results are similar to what were obtained in~\citep{Laha2020, Munoz2016}. The solid lines and dashed lines represent the constraints on $f_{\rm PBH}$ from $10^4$ FRBs with CRD and SRD, respectively. Given that the temporal width of the bursts varies and the presence of sub-bursts, we estimate that the wide range of $\overline{\Delta t}$ will represent a satisfactory projection into the potential constraints that a future survey like CHIME can achieve. First, we assume the average widths of FRBs as follows: $\overline{\Delta t}=0.1$, $ 0.3$, $1$, and $3\,\rm ms$. Moreover, we also consider a more realistic case by assuming that the distribution of widths (DoW) of future FRBs will be consistent with the one of current available FRBs. As shown in Fig.~\ref{fig3}, a smaller value of $\overline{\Delta t}$ allows one to probe lower lens masses in the $M_{\rm PBH}-f_{\rm PBH}$ plane. By using $10^4$ FRBs, the abundance of PBH $f_{\rm PBH}$ at large mass can be asymptotically constrained to $\sim0.7\%$ and $\sim0.5\%$ assuming the CRD and the SRD, respectively. In addition, compared to the CRD, constraints from FRBs with redshifts satisfying the SRD are slightly more stringent because SRD generally predicts more high-redshift FRBs.

\begin{figure*}
\centering
\subfigure[]{\includegraphics[width=0.45\textwidth, height=0.3\textwidth]{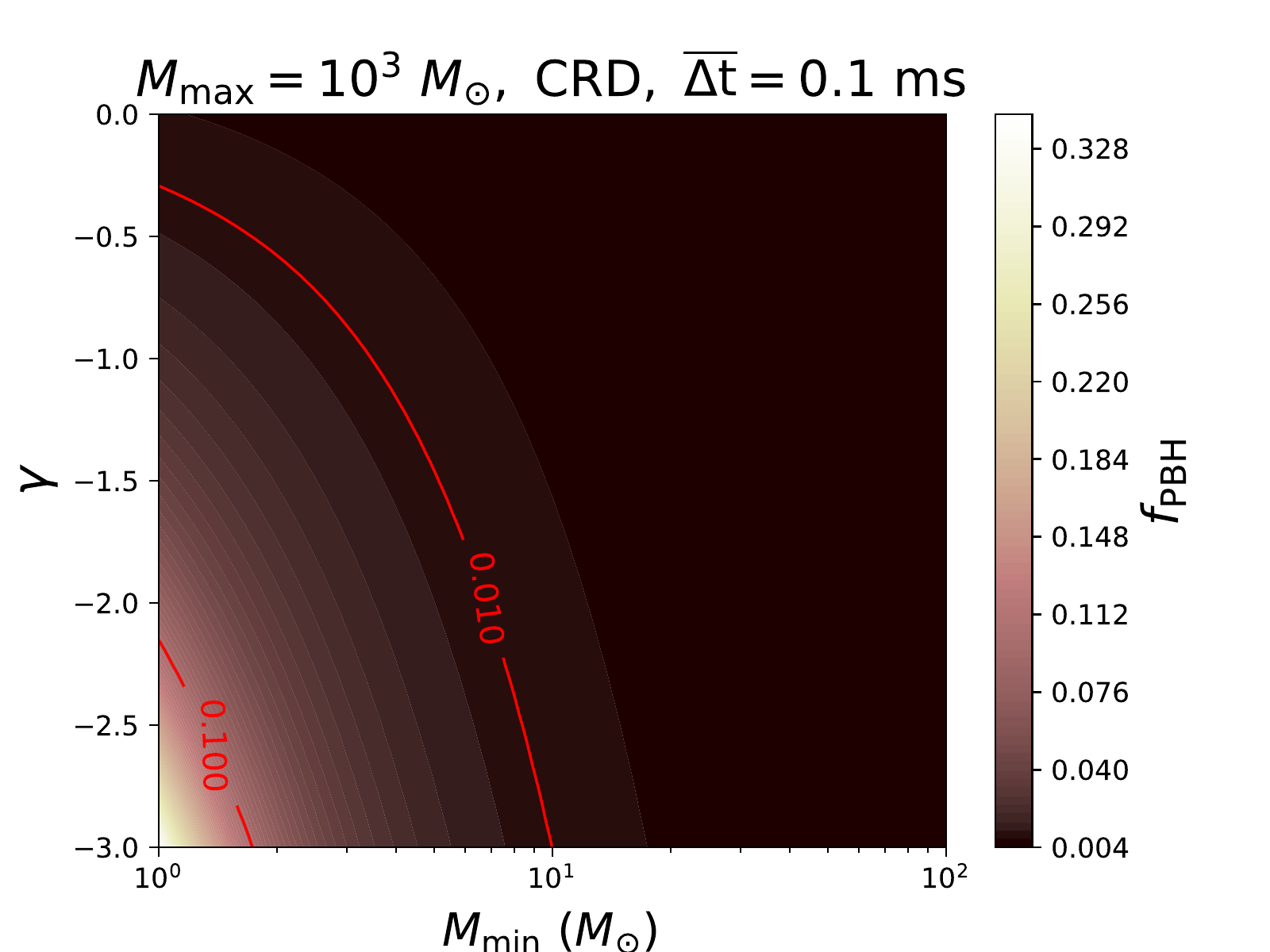}}
\subfigure[]{\includegraphics[width=0.45\textwidth, height=0.3\textwidth]{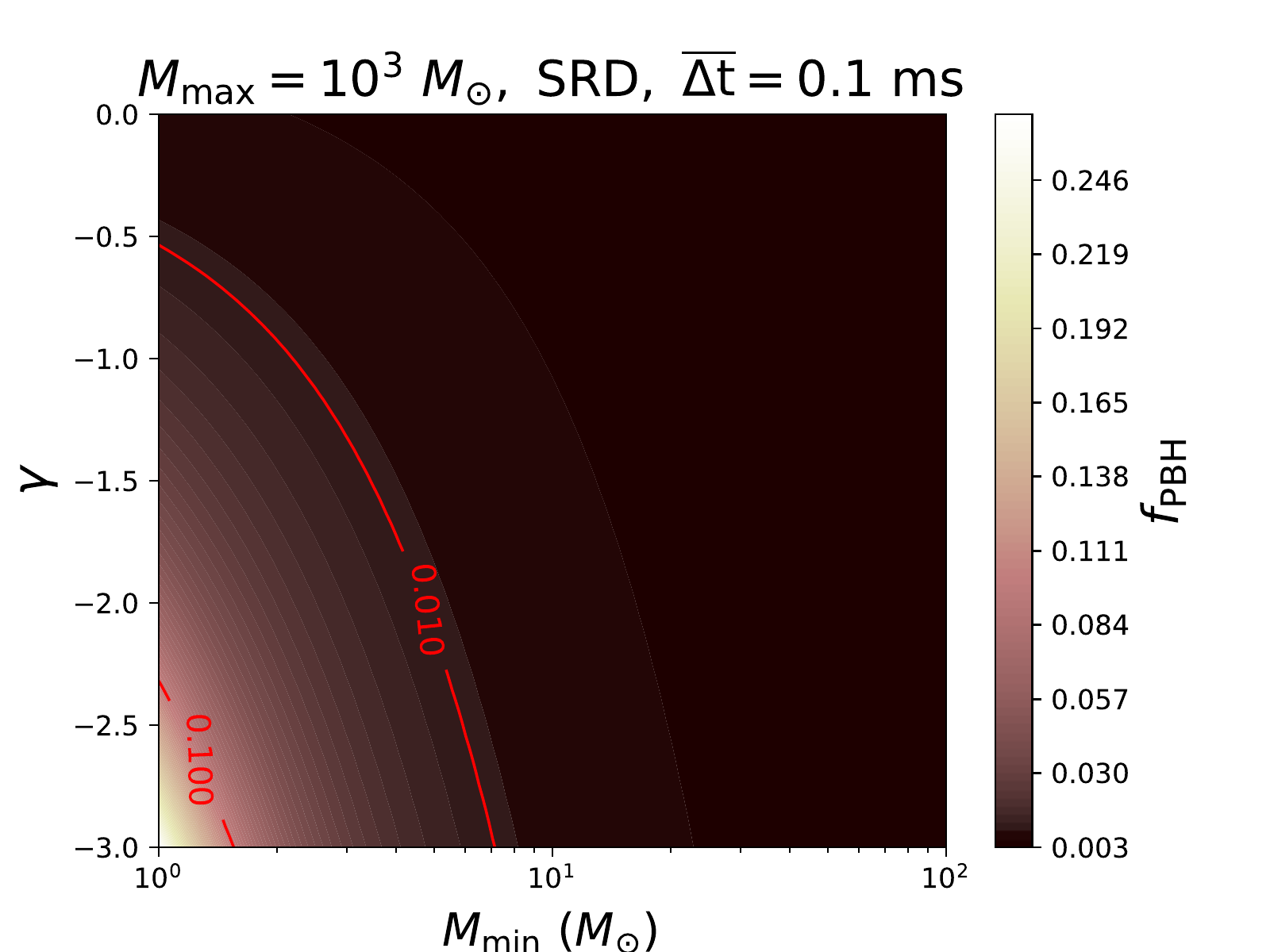}}
\subfigure[]{\includegraphics[width=0.45\textwidth, height=0.3\textwidth]{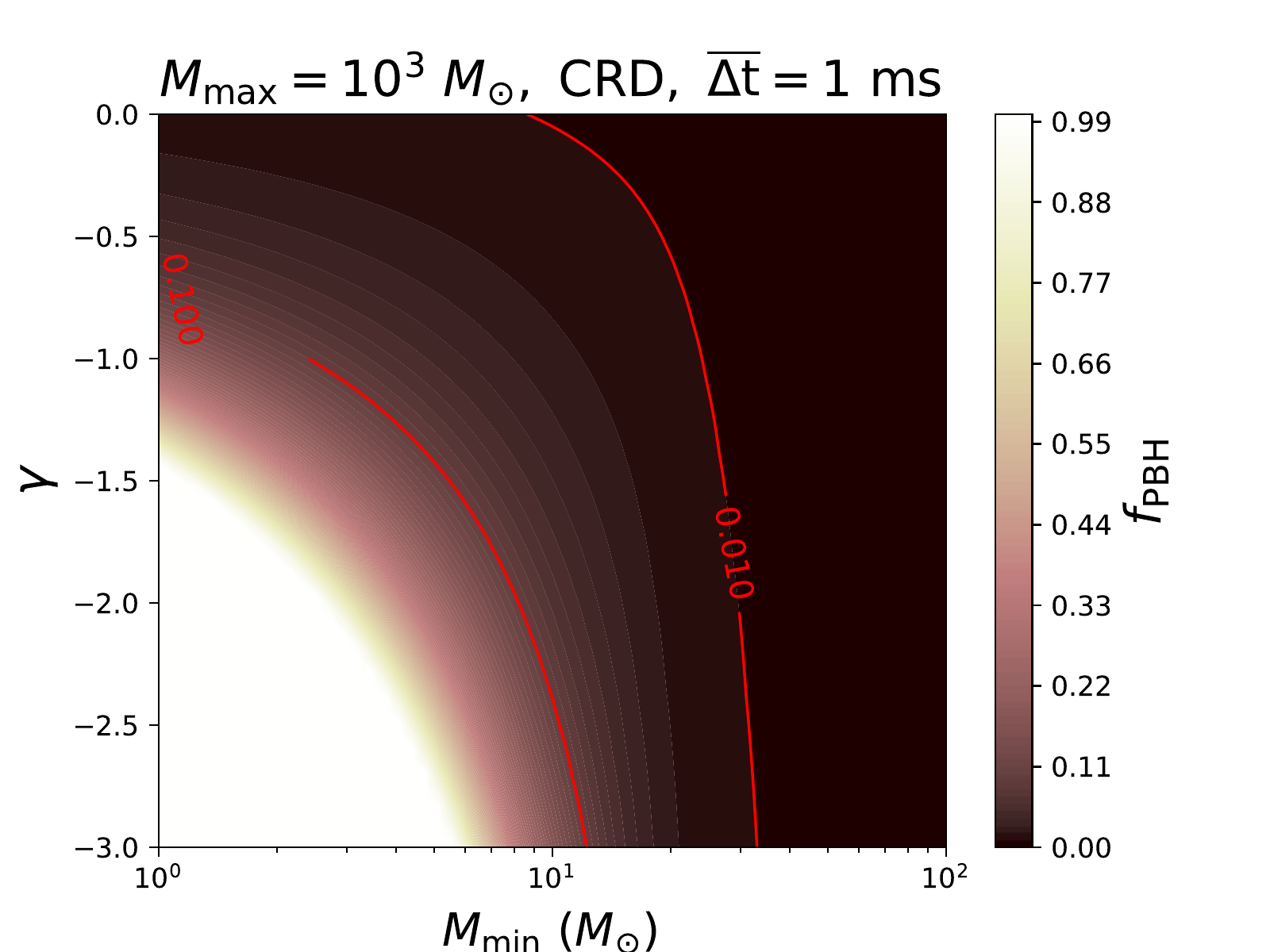}}
\subfigure[]{\includegraphics[width=0.45\textwidth, height=0.3\textwidth]{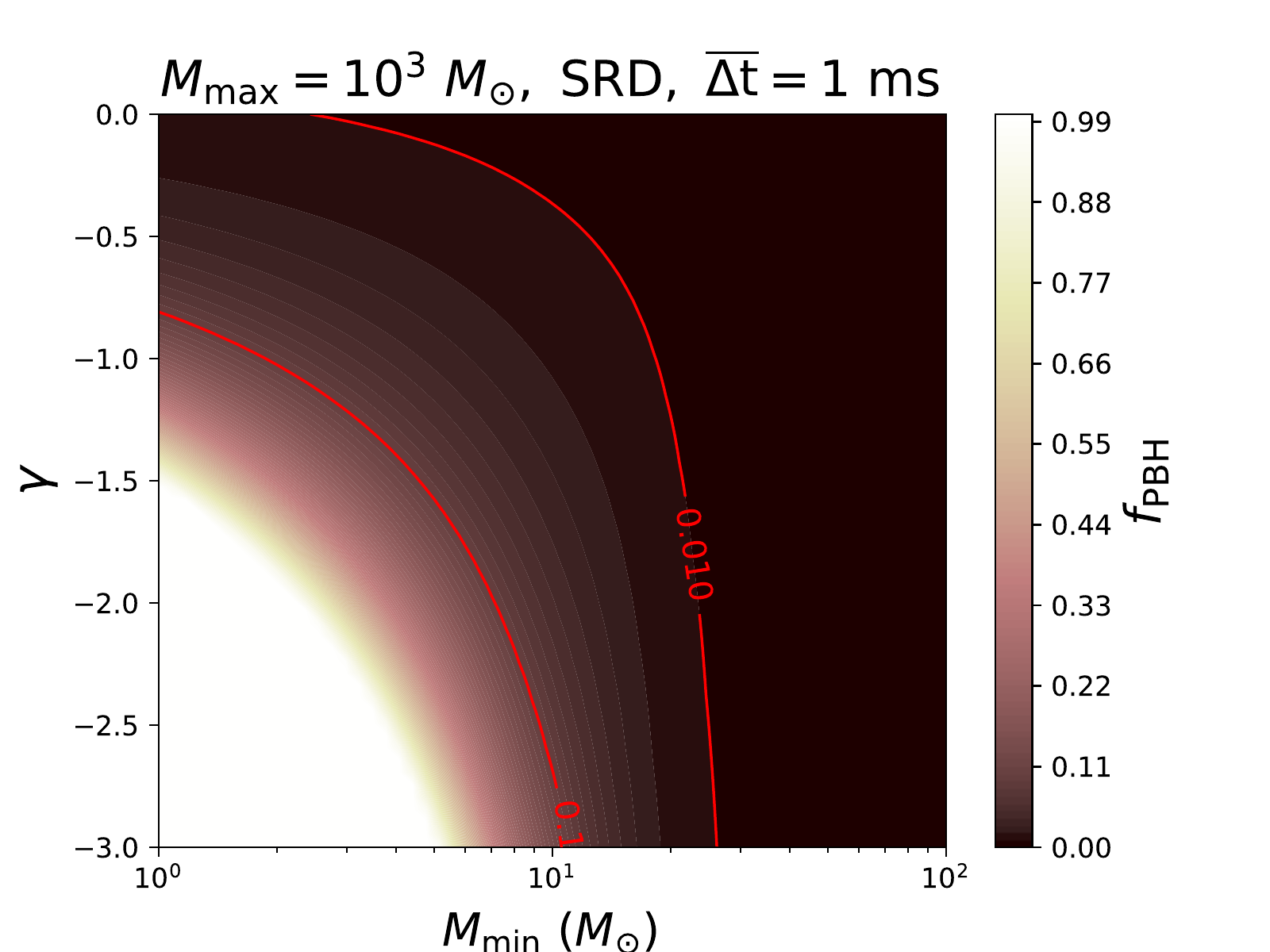}}
\caption{Constraints on the extend power-law mass function with a fixed maximum mass $M_{\max}=10^3~M_{\odot}$ from a null search result of lensing echoes in $10^4$ FRBs. In addition, we assume that $M_{\min}$ ranges from 1 $M_{\odot}$ to 100  $M_{\odot}$ and $\gamma$ is from -3 to 0. {\bf Upper left:} Results correspond to the assumptions that FRBs follow the constant-density redshift distribution and have average value of the critical time $\overline{\Delta t}=0.1~\rm ms$. {\bf Upper right:} Same as the upper left panel but for the assumption that FRBs follow the star-formation redshift distribution. {\bf Lower left:} Same as the upper left panel but for the assumption that FRBs have average value of the critical time $\overline{\Delta t}=1~\rm ms$. {\bf Lower right:} Same as the upper left panel but for the assumptions that FRBs follow the star-formation redshift distribution and have average value of the critical time $\overline{\Delta t}=1~\rm ms$.}\label{fig4}
\end{figure*}

\begin{figure*}
\centering
\subfigure[]{\includegraphics[width=0.45\textwidth, height=0.3\textwidth]{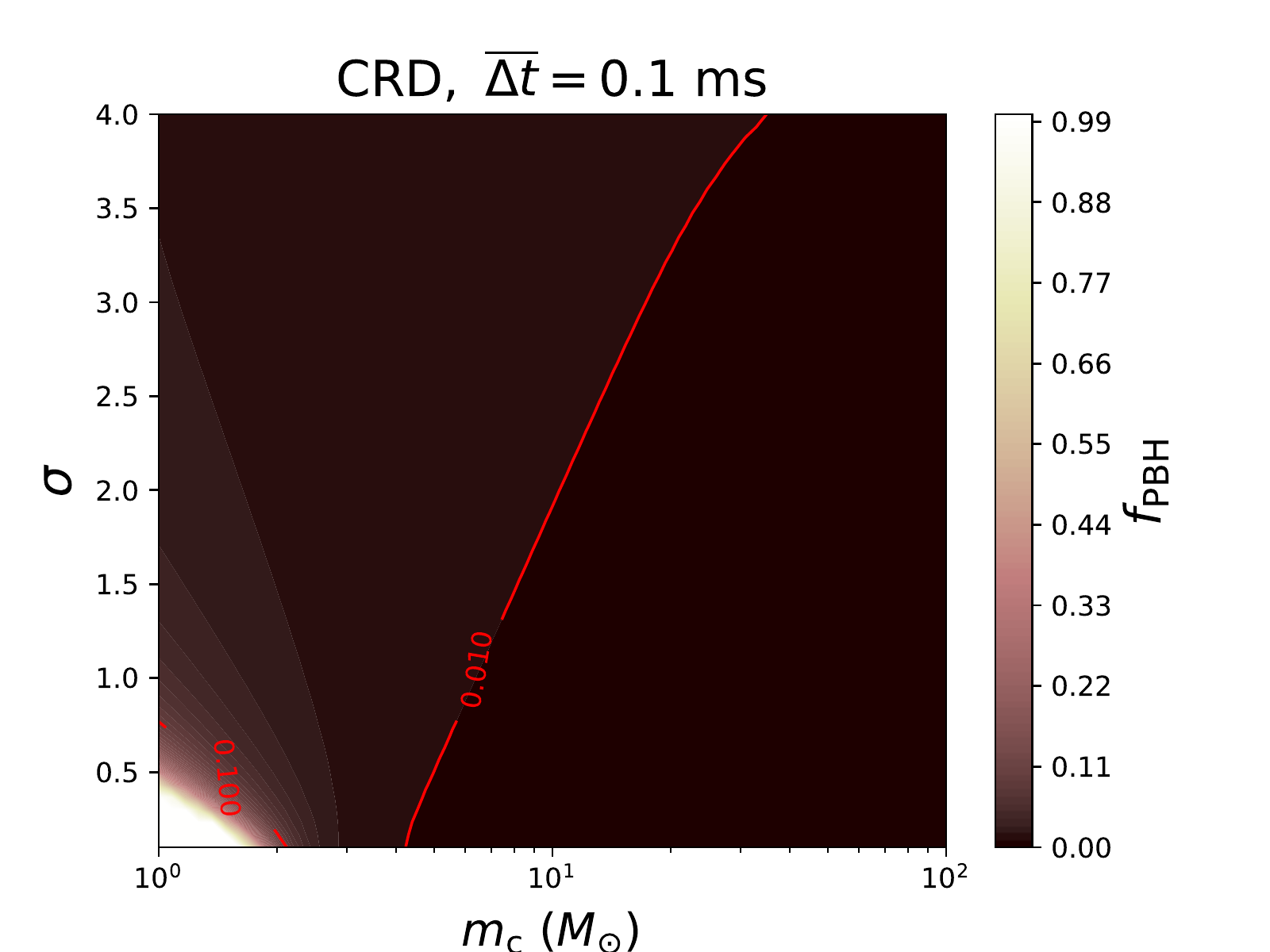}}
\subfigure[]{\includegraphics[width=0.45\textwidth, height=0.3\textwidth]{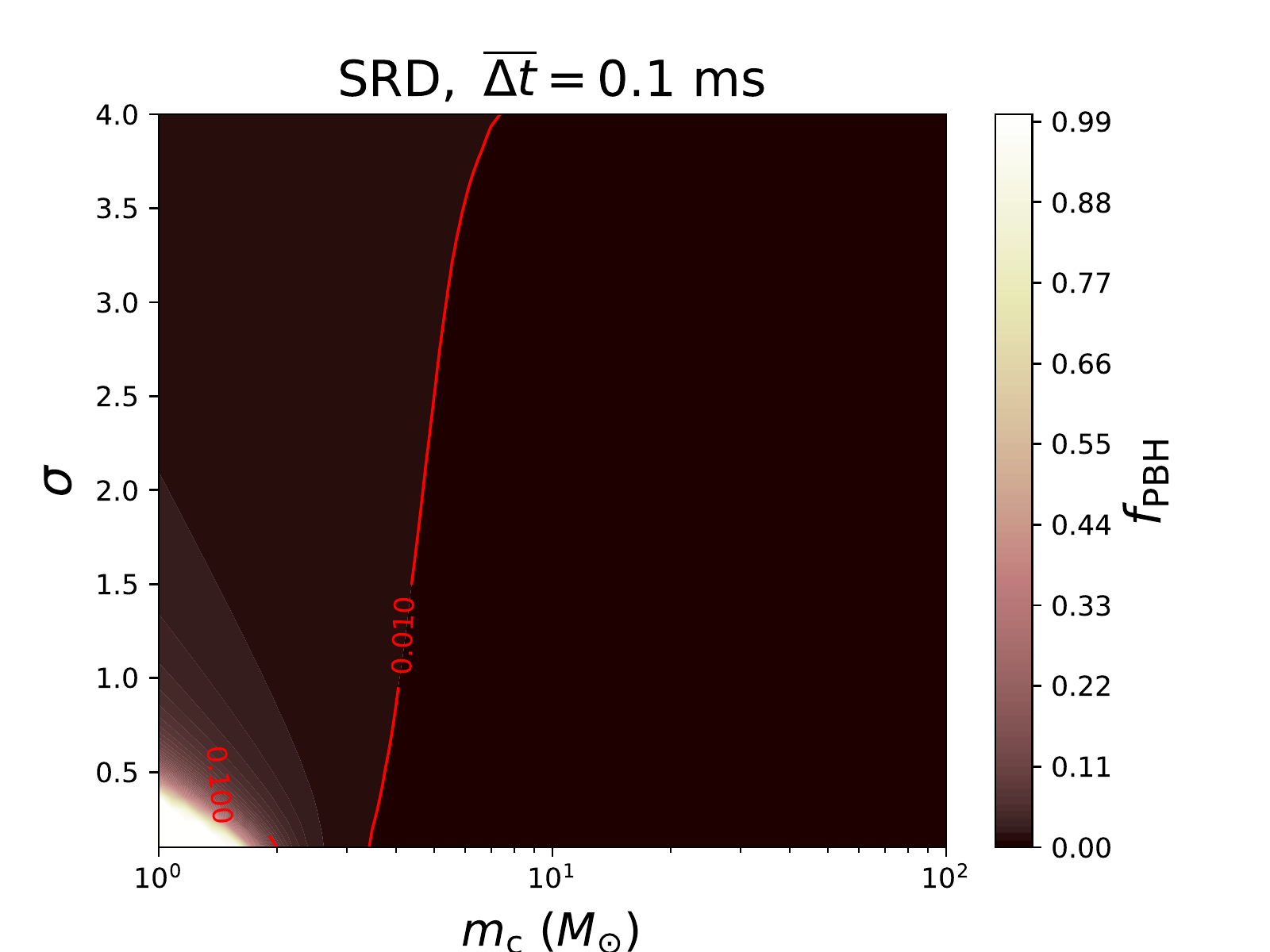}}
\subfigure[]{\includegraphics[width=0.45\textwidth, height=0.3\textwidth]{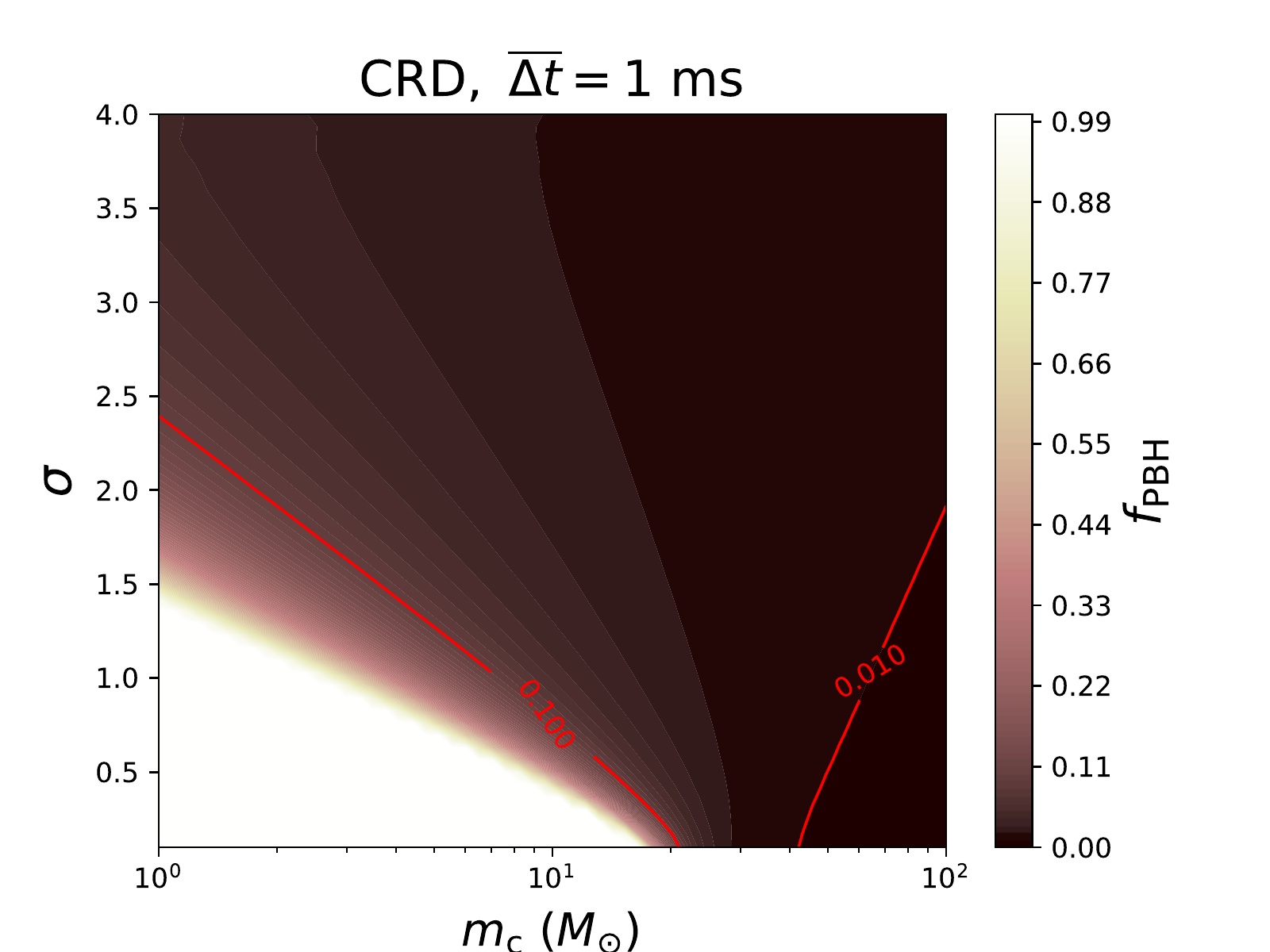}}
\subfigure[]{\includegraphics[width=0.45\textwidth, height=0.3\textwidth]{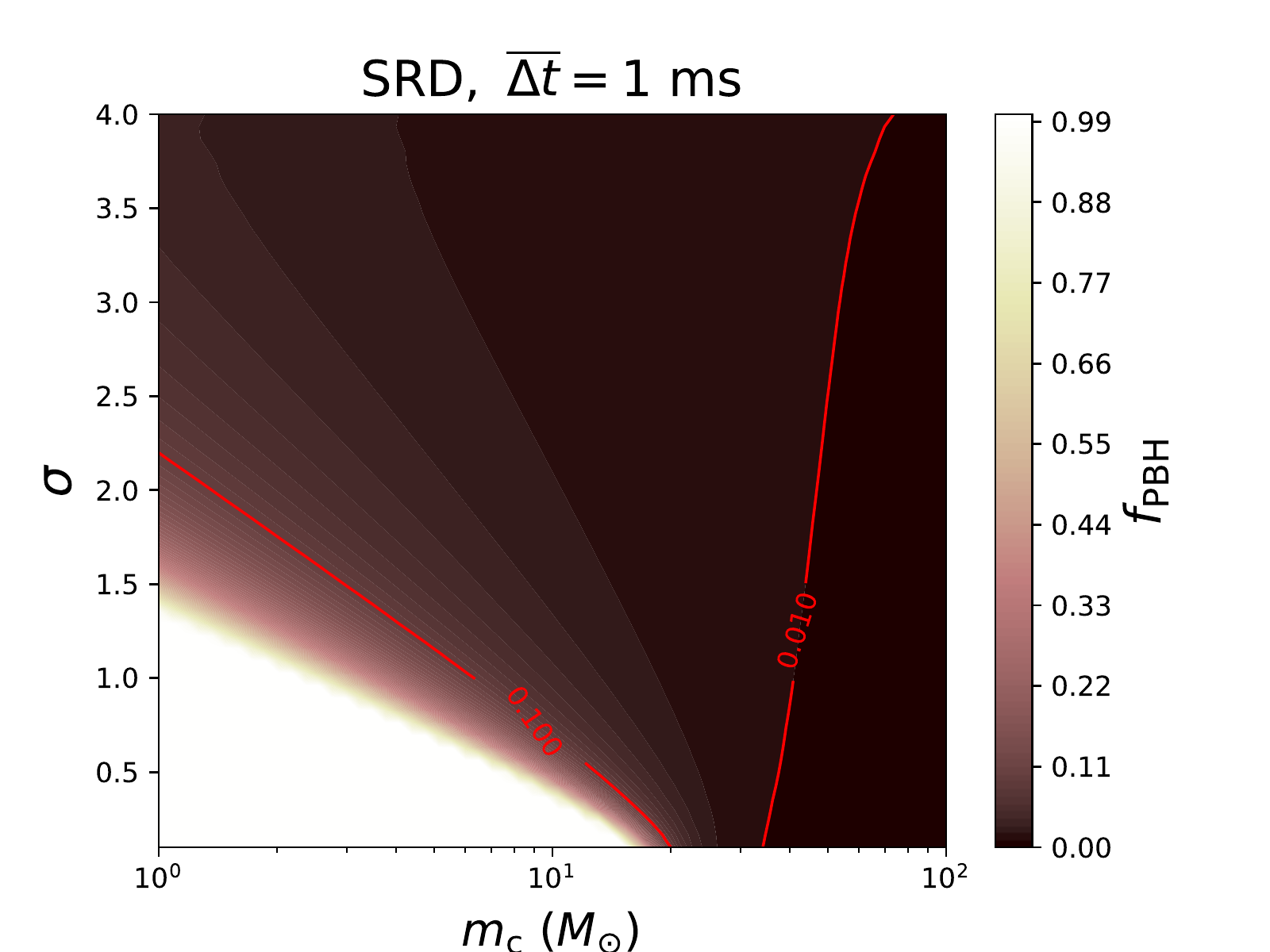}}
\caption{Same as Fig.~\ref{fig4} for the log-normal mass function with two parameters ($\sigma, m_{\rm c}$). In addition, we assume that $m_{\rm c}$ ranges from 1 $M_{\odot}$ to 100 $M_{\odot}$ and $\sigma$ is from 0.1 to 4.0, which are roughly consistent with the estimation derived from GW observations.}\label{fig5}
\end{figure*}

\begin{figure*}
\centering
\subfigure[]{\includegraphics[width=0.45\textwidth, height=0.3\textwidth]{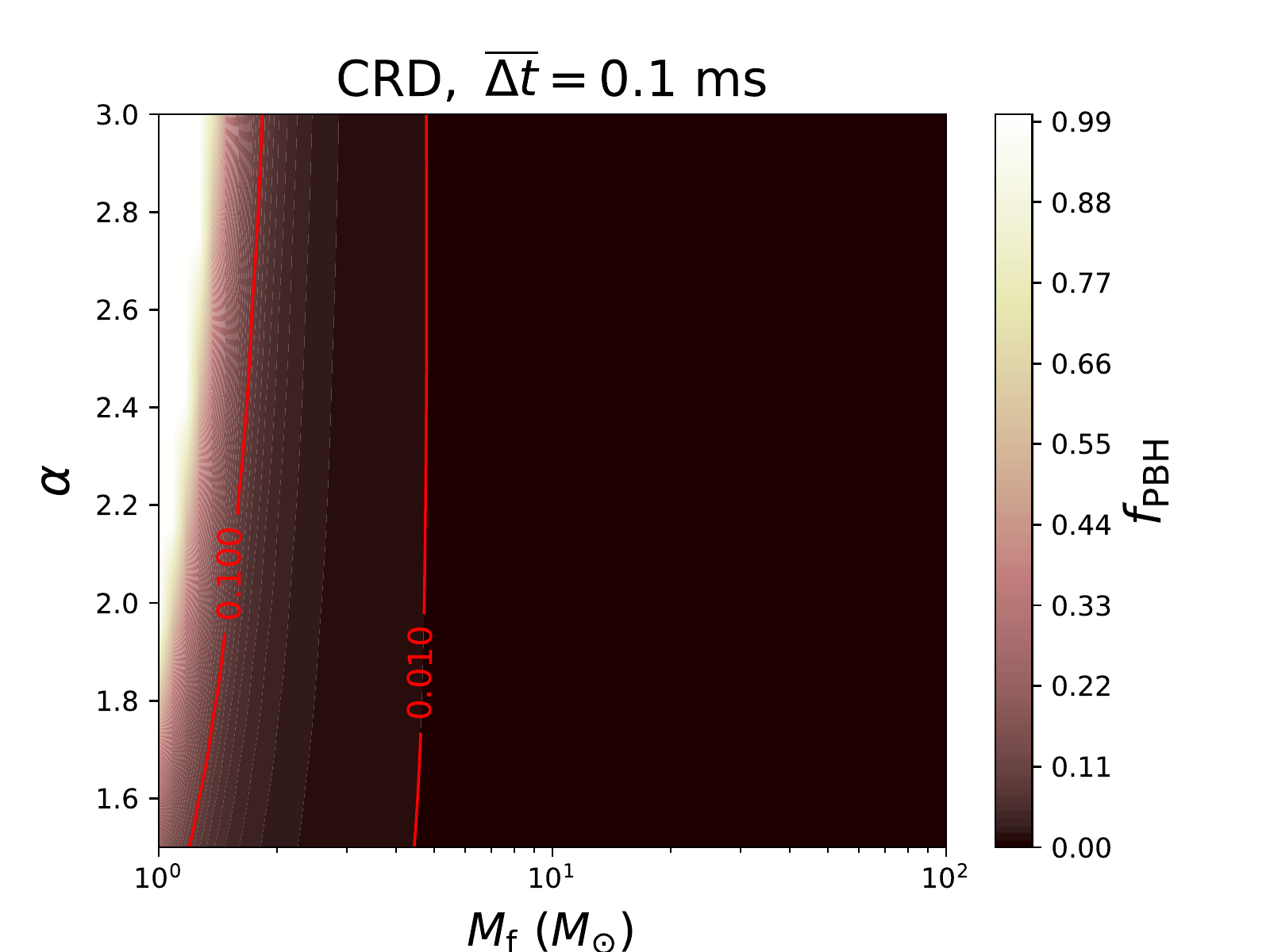}}
\subfigure[]{\includegraphics[width=0.45\textwidth, height=0.3\textwidth]{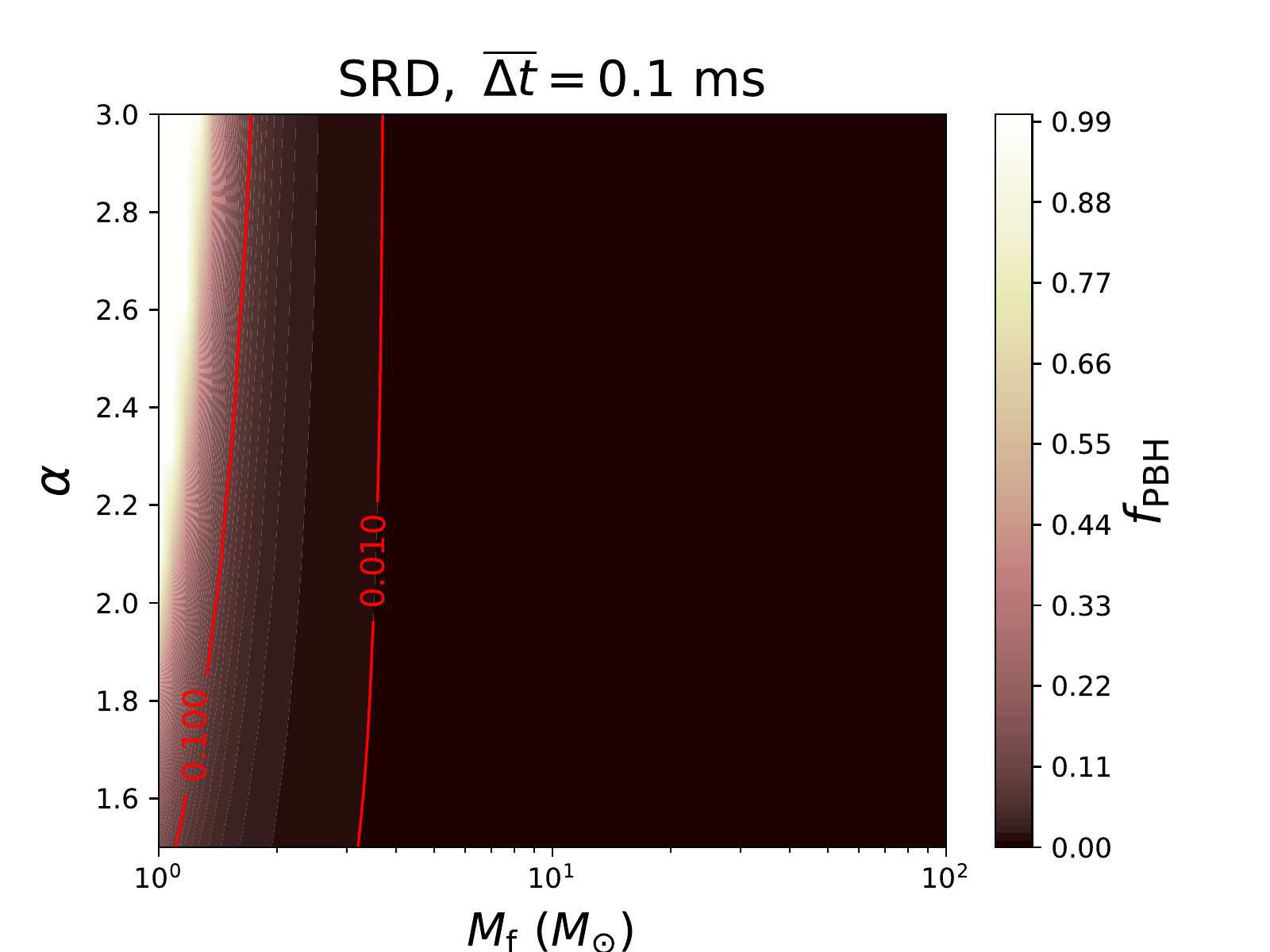}}
\subfigure[]{\includegraphics[width=0.45\textwidth, height=0.3\textwidth]{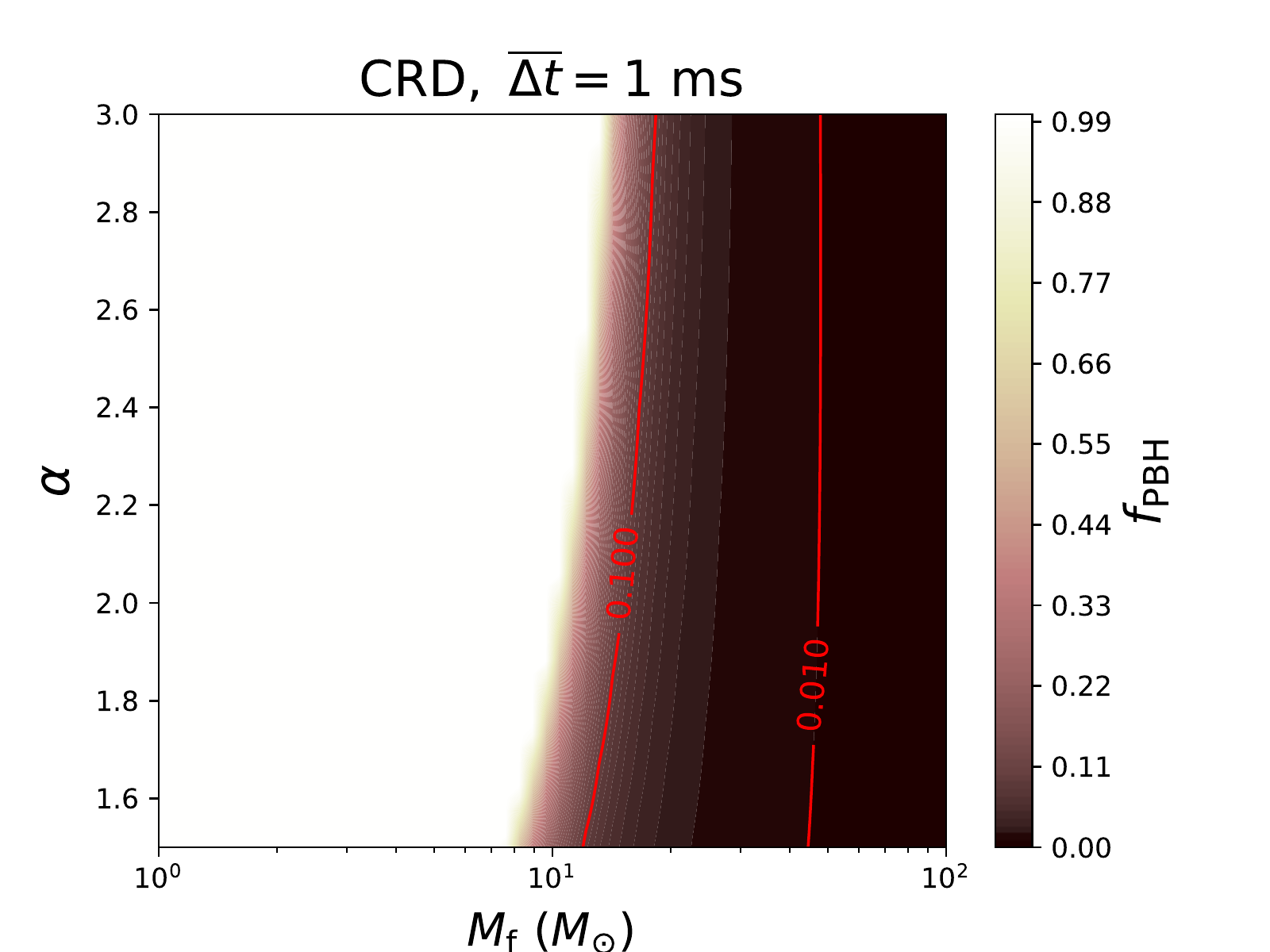}}
\subfigure[]{\includegraphics[width=0.45\textwidth, height=0.3\textwidth]{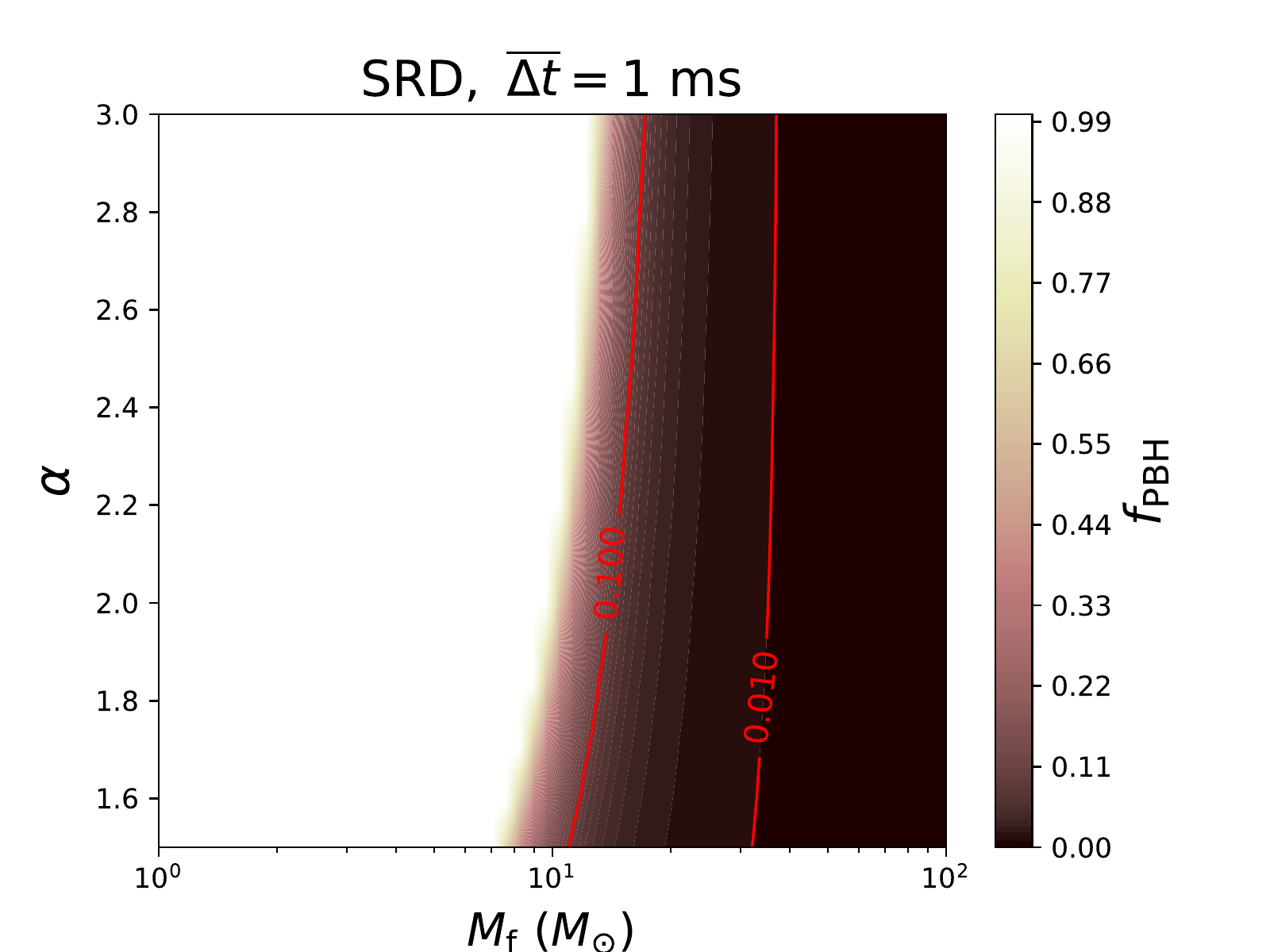}}
\caption{Same as Fig.~\ref{fig4} but for the critical collapse mass function with two parameters ($\alpha, M_{\rm f}$). In addition, we assume that $M_{\rm f}$ ranges from 1 $M_{\odot}$ to 100 $M_{\odot}$ and $\alpha$ is from 1.5 to 3.}\label{fig6}
\end{figure*}

Next, we derive the projected constraints on $f_{\rm PBH}$ with $N_{\rm FRB}=10^4$ assuming that mass of PBHs follow three different distributions. Moreover, we also assume two different redshift distributions for FRBs (CRD and SRD) and two representative average widths of them ($\overline{\Delta t}=0.1$, and $1~\rm ms$). Constraints on different parameter spaces are shown in Figs.~(\ref{fig4}, \ref{fig5}, \ref{fig6}). In addition, all the white regions in the Figs from the forecast with EMDs represent that $f_{\rm PBH}$ is more than 1. In Fig.~\ref{fig4}, we assume that the maximum value of the extended power-law mass function $M_{\max}=10^3~M_{\odot}$ and vary the minimum value $M_{\min}$ from $1~M_{\odot}$ to $100~M_{\odot}$. The regions where $10\%$ and $1\%$ of dark matter consists of PBHs are denoted in Fig.~\ref{fig4} by the red solid lines. The dependence of the results on the power-law slope $\gamma$ is similar to the real-data case, and $\overline{\Delta t}$ and SRD/CRD of FRBs all affect the results in the same way as in the MMD case. 

The forecasting constraints for the log-normal mass function are shown in Fig.~\ref{fig5}. We assume that the value of $m_{\rm c}$ varies from $1~M_{\odot}$ to $100~M_{\odot}$ and $\sigma$ is greater than $0.1$. The regions where $10\%$ and $1\%$ of dark matter can consist of PBHs are denoted by the red solid lines. Unlike the real-data case, here we probe a lower mass range. The quantitative competition between the two above-mentioned effects of broadening the mass distribution, and hence the dependence of the $f_{\rm PBH}$ constraint on $\sigma$, is more complicated. 

For the critical collapse mass function, the constraints with same $\overline{\Delta t}$ and redshift distributions of FRBs are shown in Fig.~\ref{fig6}. The contour lines where $10\%$ and $1\%$ of dark matter from PBHs are marked by the red solid lines. Similar to the real-data case, the constraint on $f_{\rm PBH}$ is sensitive to $M_f$, but insensitive  to $\alpha$.

\subsection{Comparisons with GW constraints}\label{sec4.2}
Detection of GW bursts from merges of compact object binaries is one of the most promising ways to study the mass distribution of PBHs. Constraints on the PBH scenario from the GWTC-1/GWTC-2 catalog have been widely studied via the Bayesian inference method and in the EMDs framework~\citep{Chen2019a,Wu2020,Gert2020,Kaze2020,Luca2020}. Meanwhile, as suggested in this paper, the abundance of PBHs in the mass range $1-100~M_{\odot}$ also can be well constrained from the gravitational lensing effect of upcoming FRBs. Therefore, it is interesting to compare constraints on the PBH properties from GW detection with the ones obtained from FRBs observations.  This comparison will be helpful for exploring the possibility of jointly constraining PBH scenarios from newly-developing multi-messenger observations.

\begin{figure*}
\centering
\subfigure[]{\includegraphics[width=0.45\textwidth, height=0.3\textwidth]{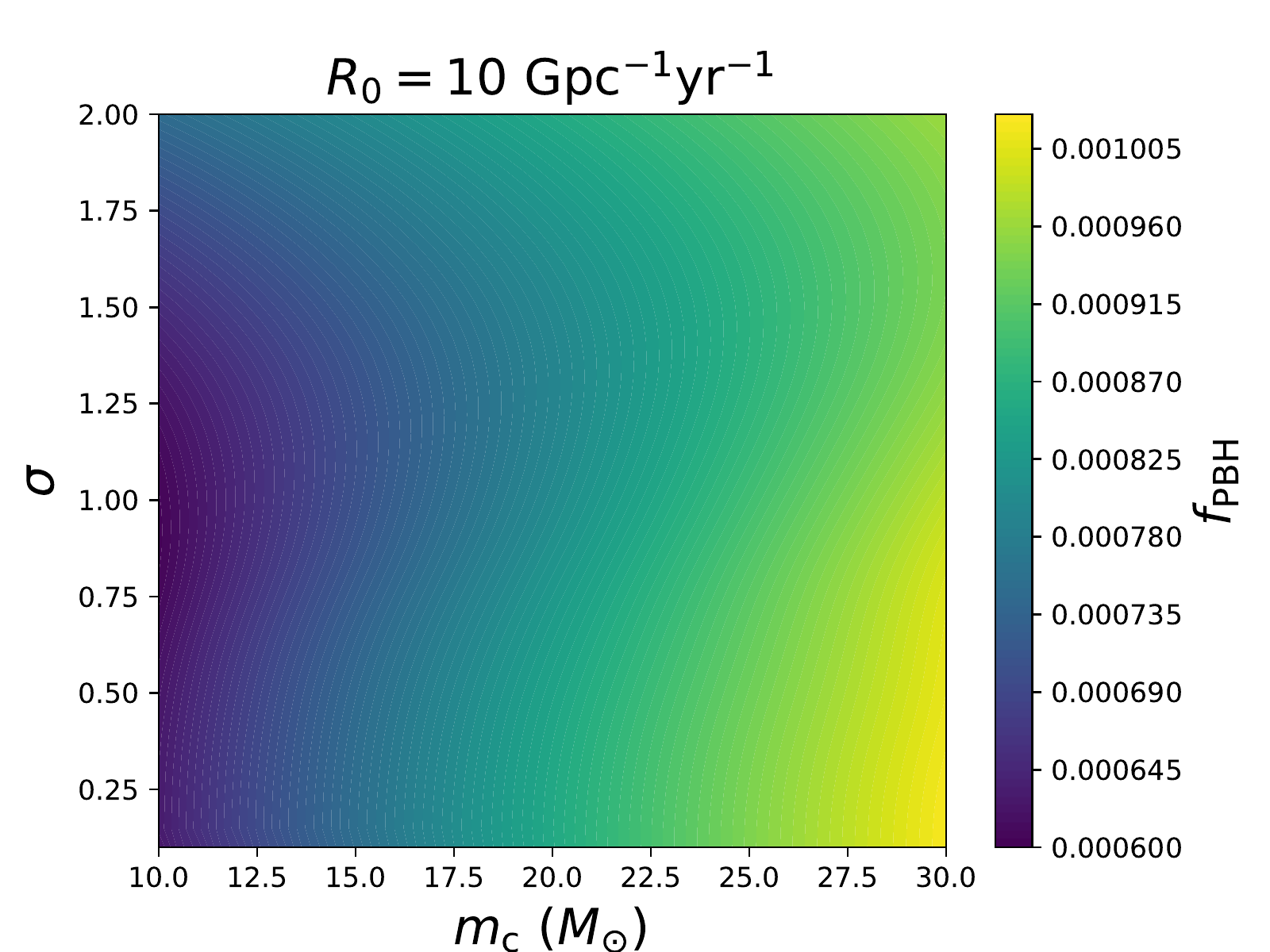}}
\subfigure[]{\includegraphics[width=0.45\textwidth, height=0.3\textwidth]{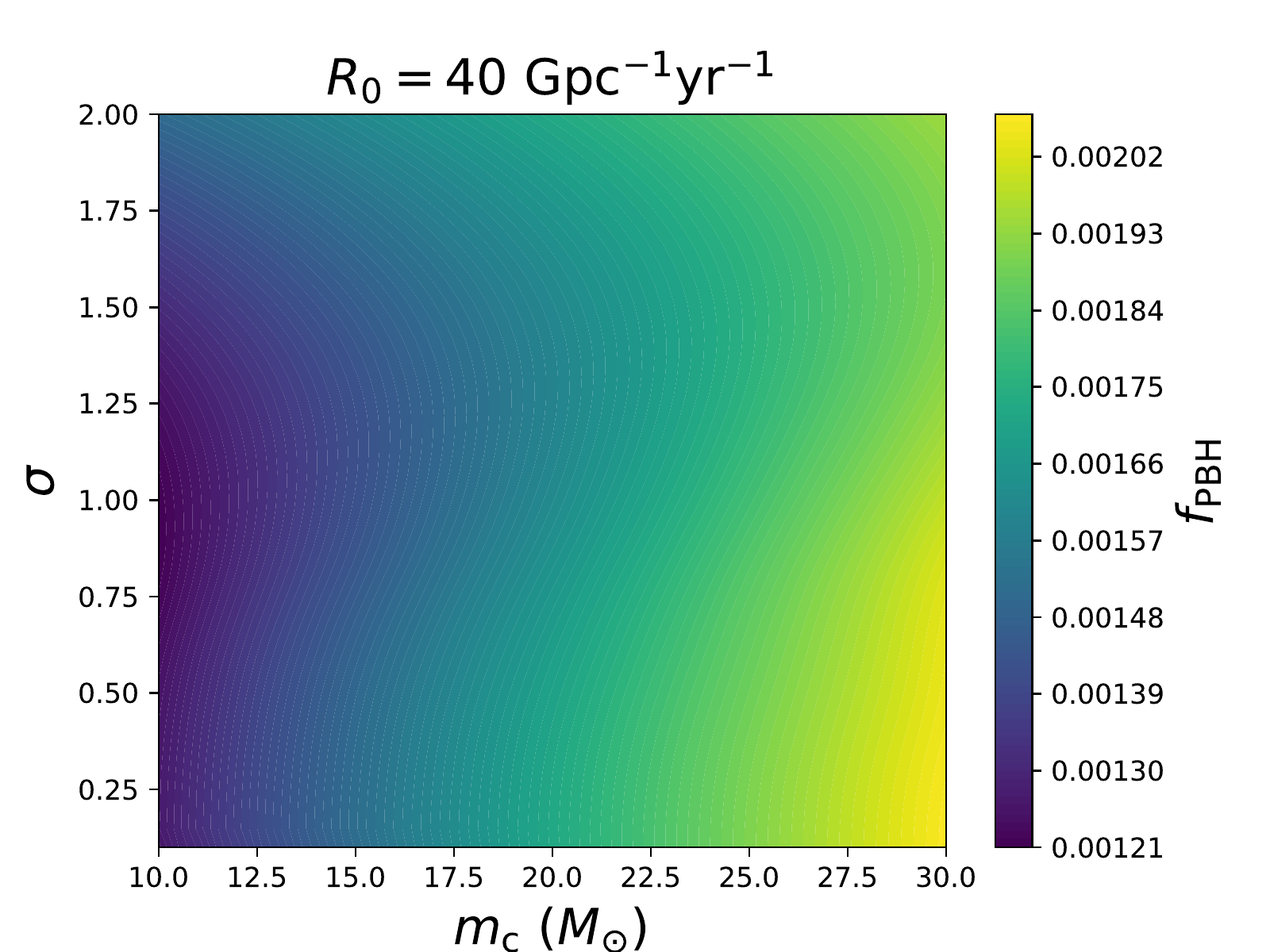}}
\subfigure[]{\includegraphics[width=0.45\textwidth, height=0.3\textwidth]{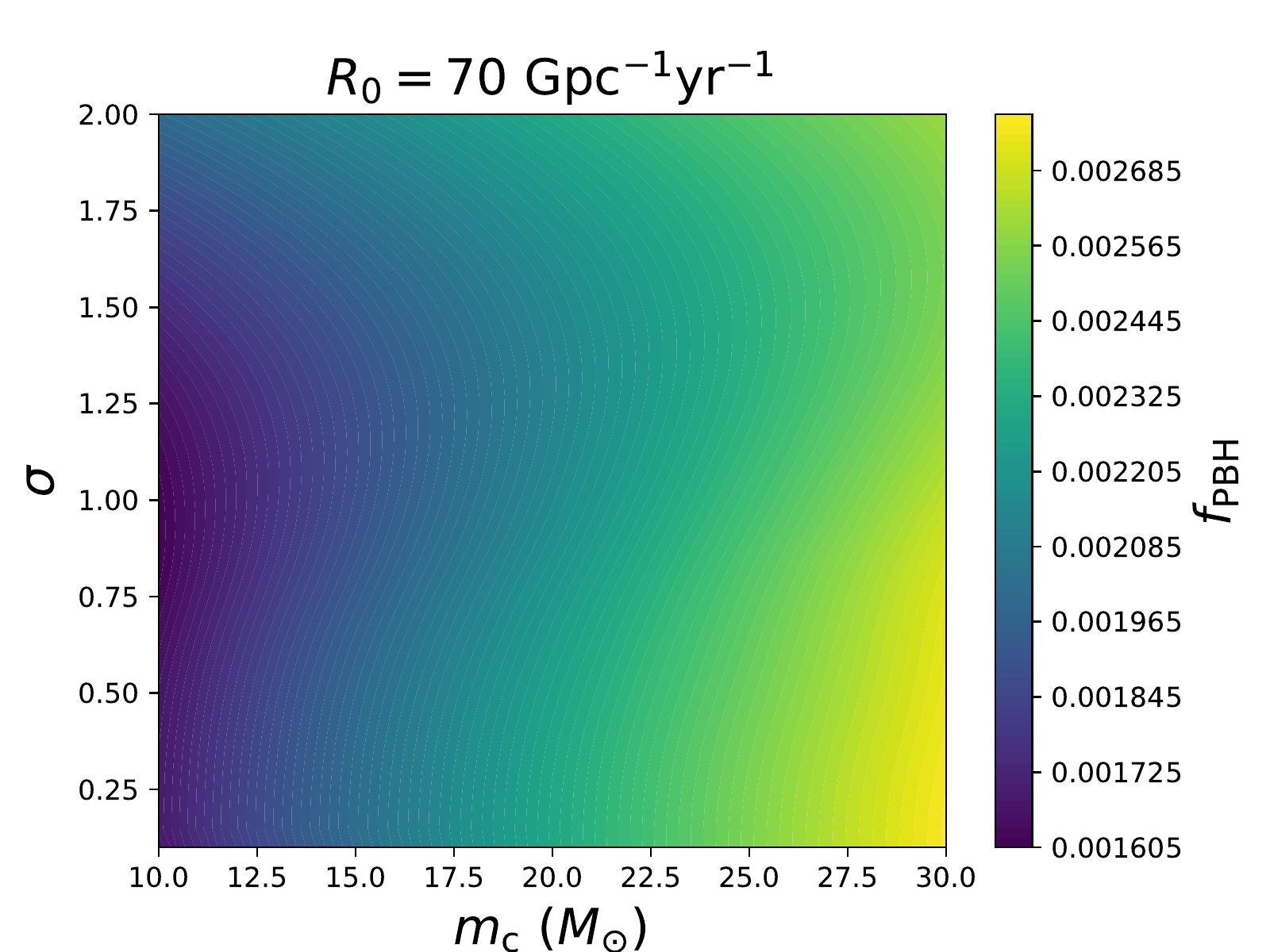}}
\subfigure[]{\includegraphics[width=0.45\textwidth, height=0.3\textwidth]{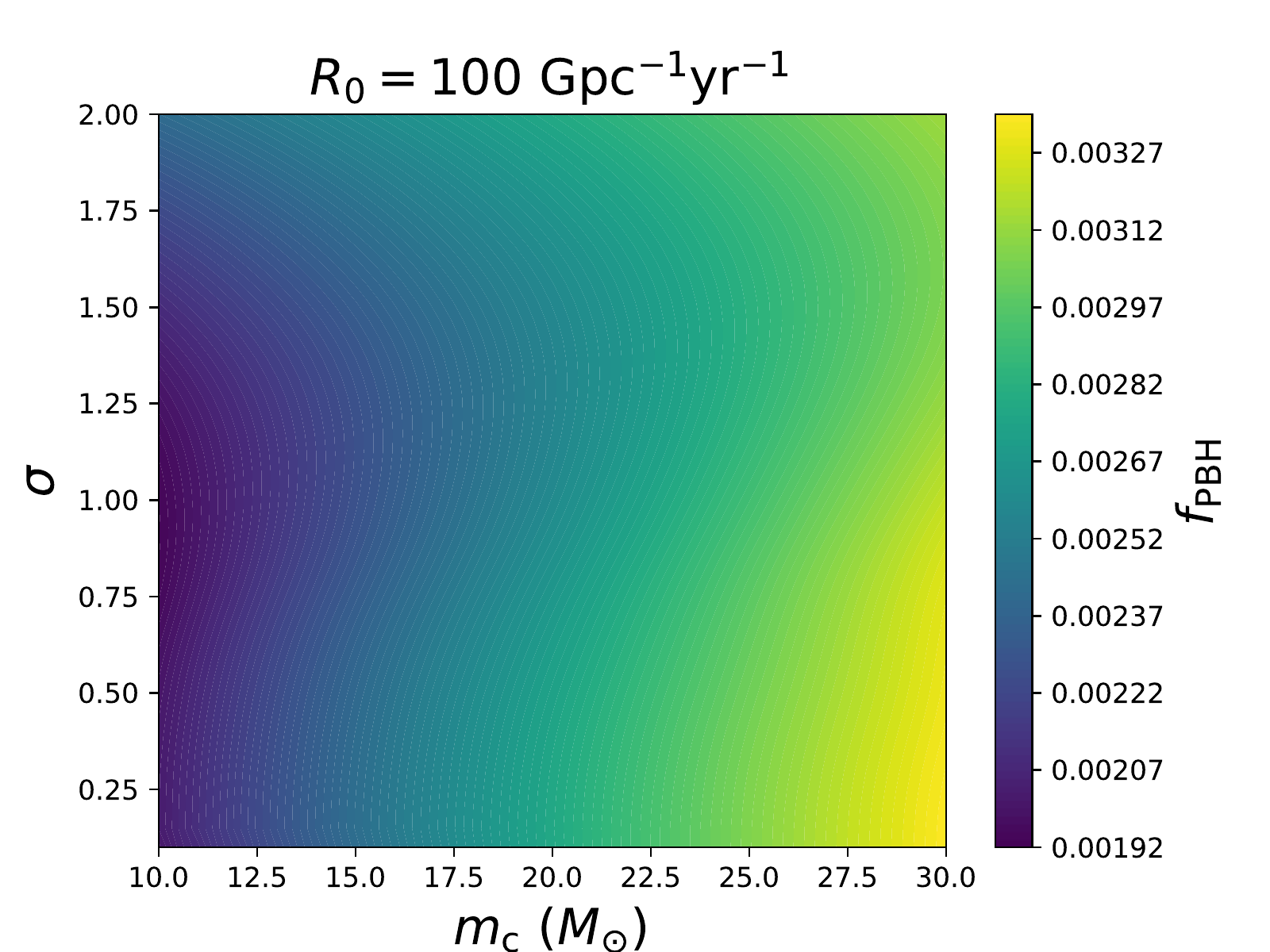}}
\caption{Constraints on the fraction of dark matter in PBHs with the log-normal mass distribution from the latest GW observations. The priors that $m_{\rm c}$ ranges from 10 $M_{\odot}$ to 30 $M_{\odot}$, and $\sigma$ is from 0.1 to 2 are considered. Panels (a, b, c, d) represent results when the merger rate of PBHs is assumed as 10, 40, 70, and $100~\rm Gpc^{-3}yr^{-1}$, respectively.}\label{fig7}
\end{figure*}

\begin{figure*}
\centering
\subfigure[]{\includegraphics[width=0.45\textwidth, height=0.3\textwidth]{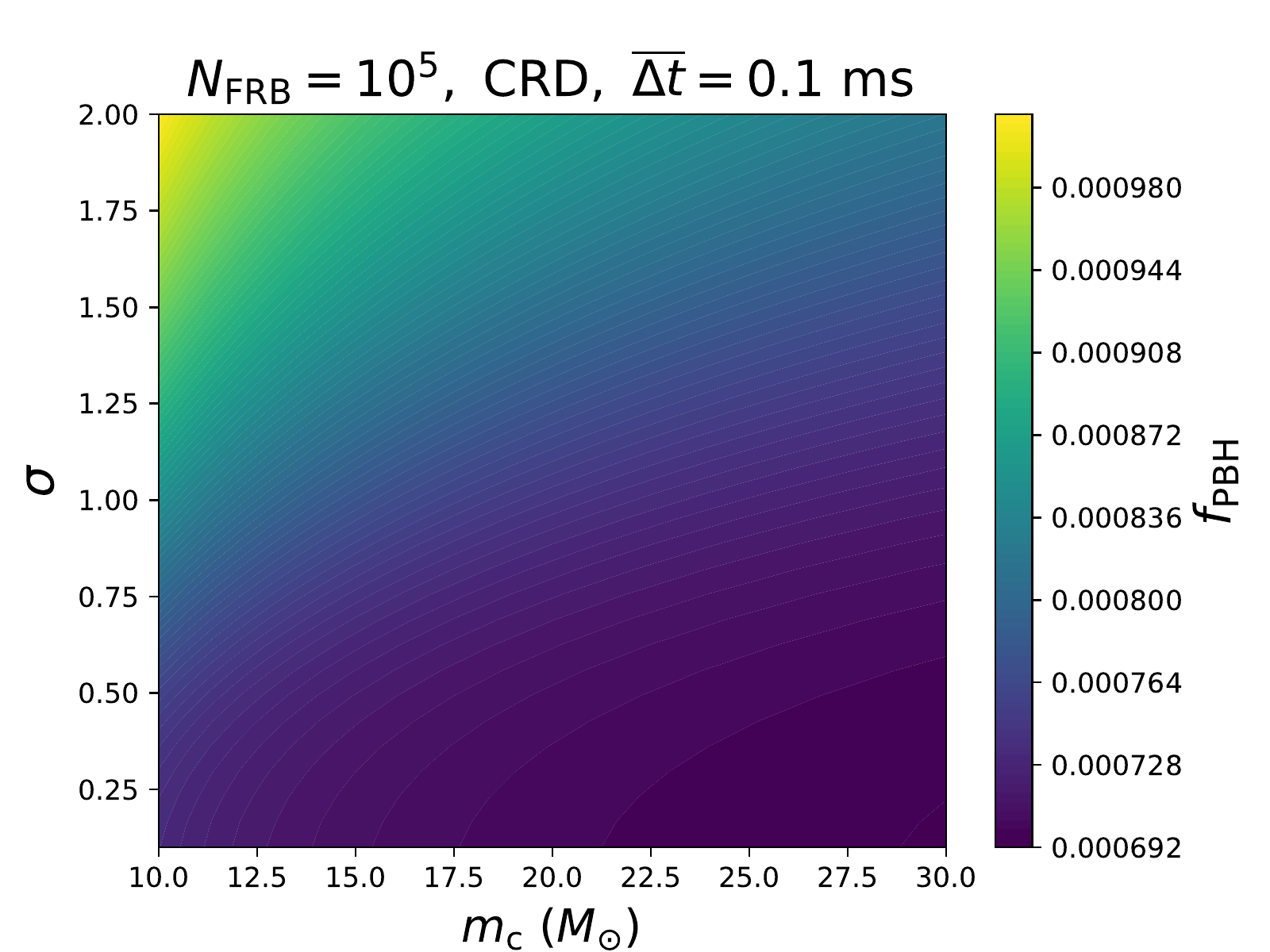}}
\subfigure[]{\includegraphics[width=0.45\textwidth, height=0.3\textwidth]{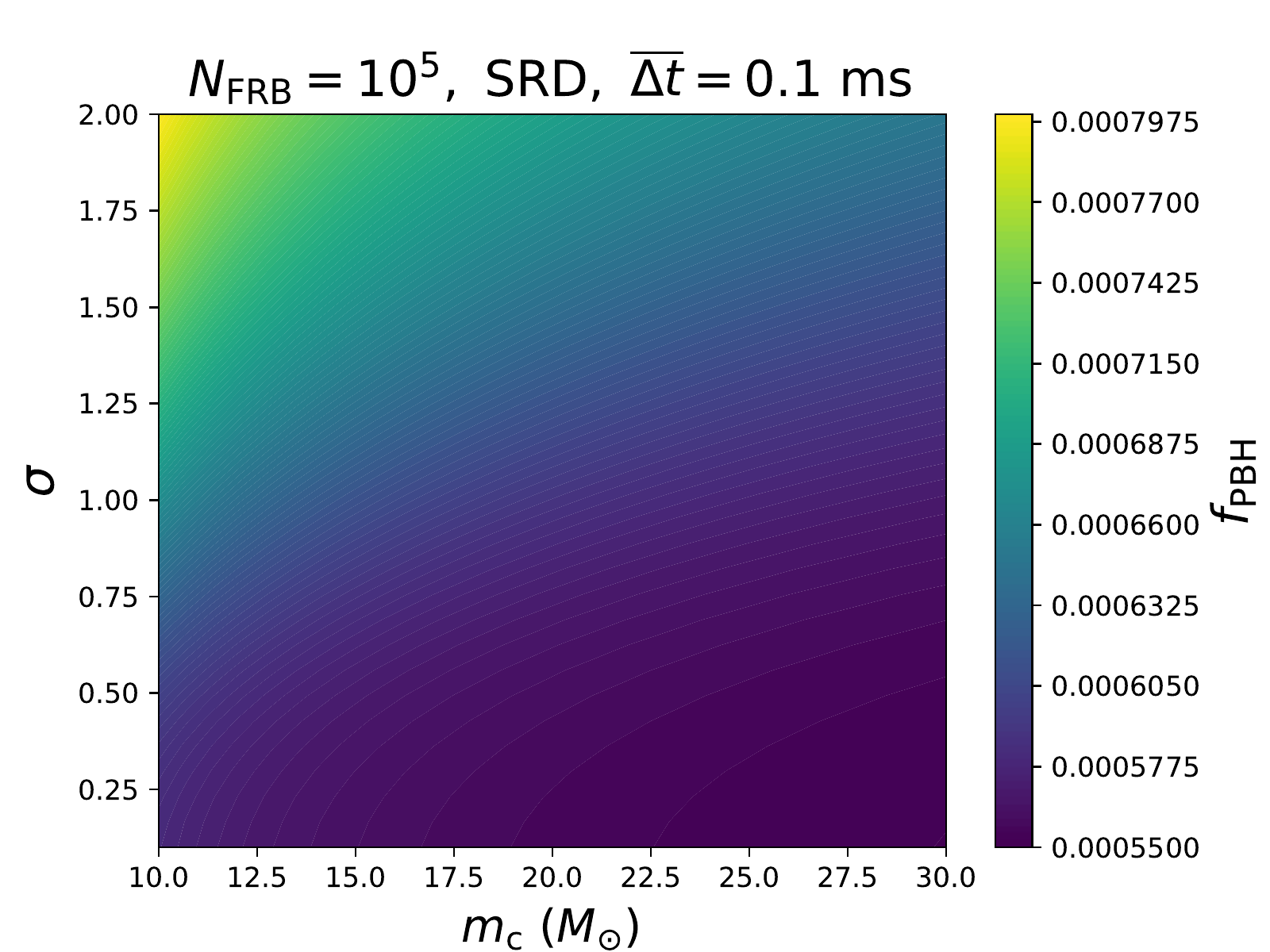}}
\subfigure[]{\includegraphics[width=0.45\textwidth, height=0.3\textwidth]{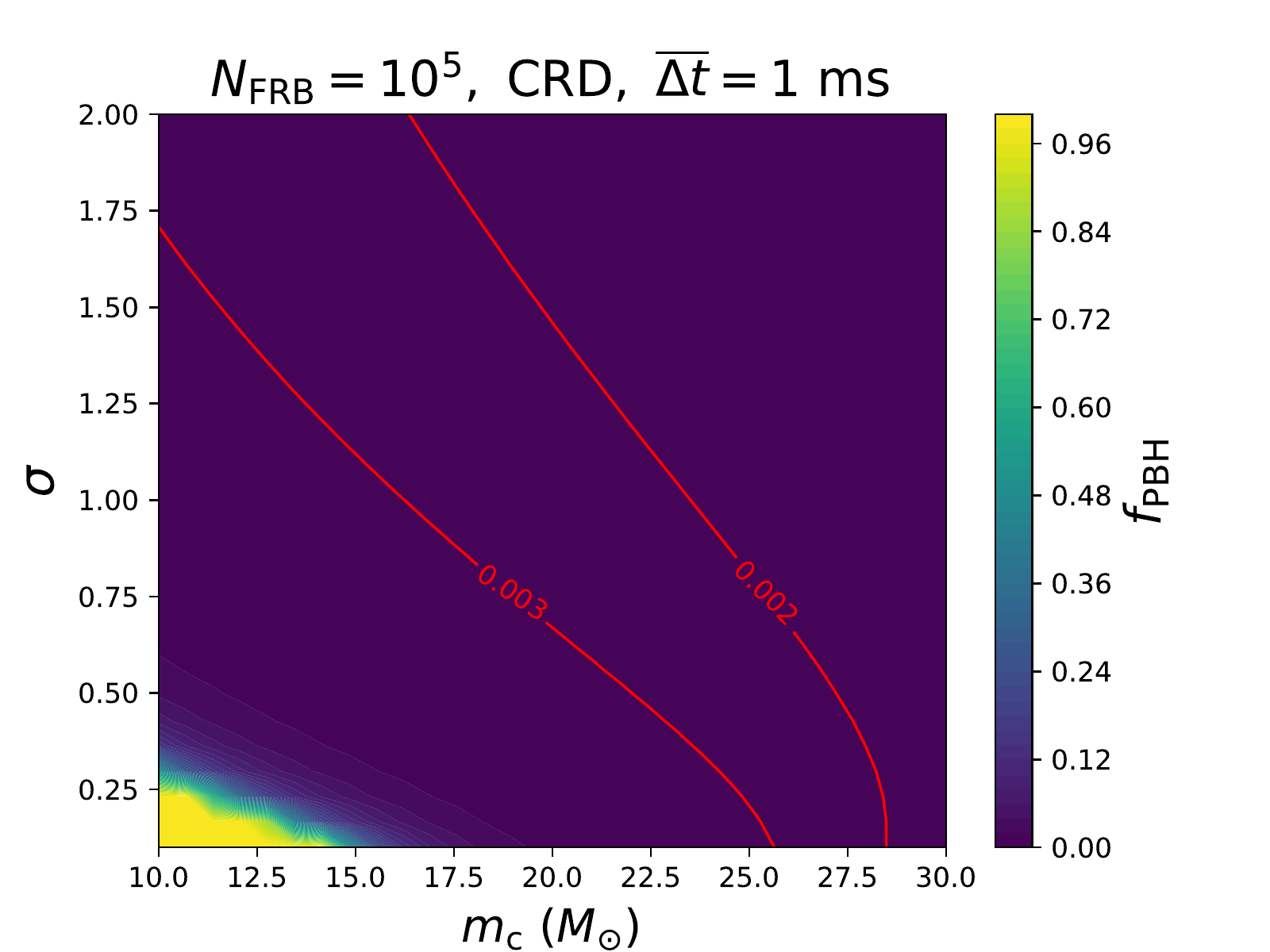}}
\subfigure[]{\includegraphics[width=0.45\textwidth, height=0.3\textwidth]{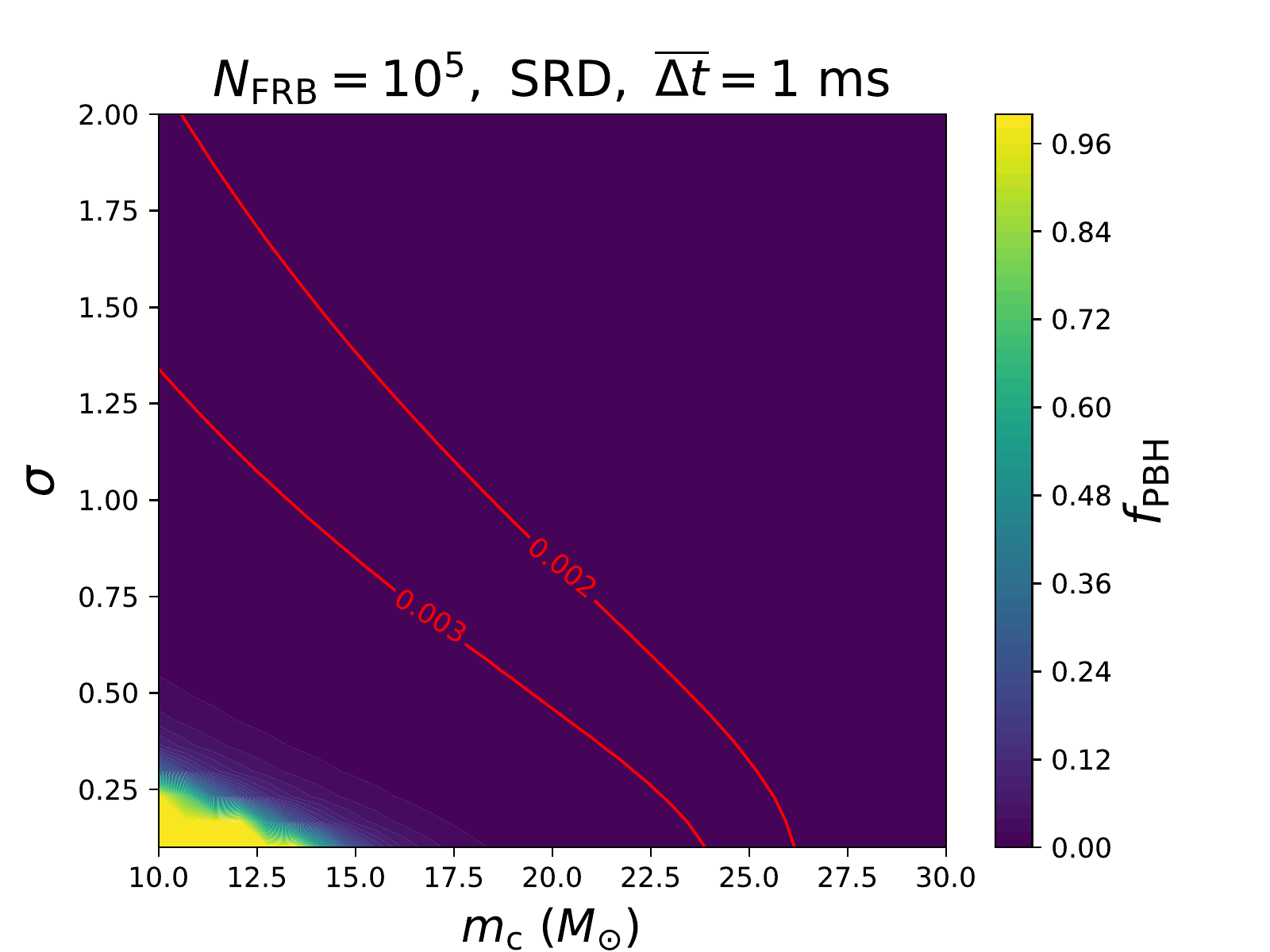}}
\caption{Same as Fig.~\ref{fig5} but for constraints based on the assumption that no lensed signal has been found in $10^5$ FRBs. In addition, we assume that $m_{\rm c}$ ranges from 10 $M_{\odot}$ to 30 $M_{\odot}$ and $\sigma$ is from 0.1 to 2.0, which are consistent with the parameter space in Fig.~\ref{fig7}}\label{fig8}
\end{figure*}

For consistency and illustration, here we apply the commonly used log-normal mass distribution to calculate the comoving merger rate density~\citep{Chen2018}
\begin{equation}\label{eq22}
\begin{split}
R_{12}(z,m_1,m_2,\sigma,m_{\rm c})=3.9\cdot10^6\bigg(\frac{t(z)}{t_0}\bigg)^{-\frac{34}{37}}f^2(f^2+\sigma_{\rm eq}^2)^{-\frac{21}{74}}
\\{\rm min}\bigg(\frac{P(m_1,\sigma,m_{\rm c})}{m_1},\frac{P(m_2,\sigma,m_{\rm c})}{m_2}\bigg)(m_1m_2)^{\frac{3}{37}}(m_1+m_2)^{\frac{36}{37}}
\\\bigg(\frac{P(m_1,\sigma,m_{\rm c})}{m_1}+\frac{P(m_2,\sigma,m_{\rm c})}{m_2}\bigg)
,
\end{split}
\end{equation}
where $t_0$ is the age of the universe, and $\sigma_{\rm eq}$ is the variance of density perturbations of the rest dark matter on scale of order $\mathscr{O}(10^0\sim10^3)~M_{\odot}$ at radiation-matter equality. Here $f$ is the total abundance of PBHs in non-relativistic matter, and relates to the fraction of PBHs via $f_{\rm PBH}\equiv\Omega_{\rm PBH}/\Omega_{\rm CDM}\approx f/0.85$. Integrating over the component masses in the merger rate density, we can obtain the merger rate as a function with respect to redshift,
\begin{equation}\label{eq23}
R(z)=\int R_{12}(z,m_1,m_2,\sigma,m_{rm c})dm_1dm_2.
\end{equation}
The local merger rate is defined as $R_0\equiv R(z=0)$. In this paper, the local merger rate of PBHs is assumed as 10, 40, 70, and $100~\rm Gpc^{-3}yr^{-1}$, respectively, to infer the fraction of dark matter in the form of PBHs. Then this inferred $f_{\rm PBH}$ will be compared with the constraints from FRB observations. As shown in Fig.~\ref{fig7}, we assume that, for the log-normal mass distribution in the merger rate Eq.~\ref{eq23}, $m_{\rm c}$ is in the range of 10-30 $M_{\odot}$ and $\sigma$ ranges from 0.1 to 2. These priors are roughly consistent with the parameter space constrained from the latest GW data. It is suggested that, for the assumption $R_0=10~\rm Gpc^{-3}yr^{-1}$, $f_{\rm PBH}$ is estimated to be $<0.1\%$, whereas for the assumption $R_0=100~\rm Gpc^{-3}yr^{-1}$, $f_{\rm PBH}$ is estimated to be $<0.3\%$. However, as shown in Fig.~\ref{fig5}, the upper limits of $f_{\rm PBH}$ is approximately constrained to be $0.5\%$ ($0.7\%$) from null search result of lensing copies in $10^4$ FRBs  when the SRD (CRD) and $\overline{\Delta t}=0.1~\rm ms$ is assumed. Obviously, constraints on the upper limit of $f_{\rm PBH}$ from the null search result of $N_{\rm FRB}=10^4$ FRBs is slightly weaker than the one estimated from GW observations. That is, the constrained regions in the log-normal mass distribution parameter space from observations of these two kind messengers are hardly to overlap. Therefore, in this paper we make a further assumption of a null search of lensing phenomenon in $10^5$ FRBs with value of the critical time $\overline{\Delta t}=0.1,$ and $1~\rm ms$ with two above-mentioned redshift distributions. As shown in Fig.~\ref{fig8}, the contour lines representing $0.3\%$ and $0.2\%$ of dark matter in the form of PBHs are denoted by the red solid lines. It is found that, there are considerable areas in the parameter space with the upper limits of $f_{\rm PBH}$ less than $0.1\%$ when $\overline{\Delta t}=0.1~\rm ms$ is assumed. In contrast, for the assumption $\overline{\Delta t}=1~\rm ms$, areas in the parameter space with the upper limits of $f_{\rm PBH}$ less than $0.1\%$ disappear. However, most of the upper limits of $f_{\rm PBH}$ in the parameter spaces are less than $0.3\%$. Generally speaking, in the parameter space, there are considerable overlap between the constraints from GW observations and a null search of lensing candidate in $10^5$ FRBs. This overlap makes it possible to get joint constraints on the properties of stellar mass PBHs from the upcoming multi-messenger observations.

\section{conclusions and discussions}\label{sec5}
The recent detection of GW from binary black hole merger have caused great interests in stellar mass PBHs. Consequently, detection of GW has been deemed as one of the most promising ways to examine the nature of PBHs. Meanwhile, in the electromagnetic domain, gravitational lensing effect of prolific transients with millisecond duration, e.g. FRBs, has been proposed as one of the cleanest probes for exploring properties of PBHs in the mass range $1-100~M_{\odot}$. In this paper, we first derive constraints on the abundance of PBH from the latest FRB observations and obtain that, for $M_{\rm PBH}\gtrsim 10~M_{\odot}$ an upper bound of $f_{\rm PBH}$ can be achieved. The $1\sigma$ bound saturates to $\sim 10\%$ in the large-mass ($\gtrsim 10^3~M_{\odot}$) end. This current constraint is weak but also of great significance in providing complementary information from observations of the newly discovered transient. Moreover, we investigate the dependence of results on the mass functions of PBH by taking three extended mass distributions, which are related to well-defined theoretical motivations, into consideration. It appears that, in general, the latest FRB observations yield to consistent constraints on the upper limit of $f_{\rm PBH}$ for different mass distribution scenarios. 

In addition to the constraints from currently available FRBs, we forecast the constraining power of upcoming FRB observations on the properties of PBHs by taking possible impact factors, such as the mass distribution function and the redshift distribution of FRBs, into account. First, we conservatively assume a sample of $10^4$ FRBs, which is probably to be detected by the upcoming wide-field radio arrays (like DSA-2000) in one year. We find that, for most cases, the upper limit of $f_{\rm PBH}$ at large mass can be asymptotically constrained to $\sim0.7\%$ from the null search of lensing event in $10^4$ FRBs. Meanwhile, we also estimate the abundance of PBHs from GW observations. Our results indicate that, compared with upcoming $10^4$ FRBs, current GW observations can present much more stringent constraints on the abundance of PBH. Therefore, we further investigate forecasting implications from $10^5$ FRBs which might be accumulated in several years and find that, in the framework of the log-normal mass distribution, the upper limit of $f_{\rm PBH}$ can be asymptotically constrained to $\sim0.1\%$. This constraint is close to the one estimated from GW observations. As a result, in the parameter space of the mass distribution, there would be significant overlap between areas constrained from FRB observations and the ones from GW detection. This merit is of great importance for the possibility of deriving joint constraints on the abundance and mass distribution of PBHs from the combination of forthcoming complementary multi-messenger observations. It is foreseen that these joint constraints will be very helpful for exploring the nature of PBHs in the stellar mass range, or even their formation mechanisms relating to the evolution theories of the early universe.

\section*{Acknowledgements}
We would like to thank Zu-Cheng Chen and You Wu for their helpful discussions about estimation for mass distribution of PBHs from GW observations. This work was supported by the National Natural Science Foundation of China under Grants Nos. 11920101003, 11722324, 11603003, 11633001,  12073088, and U1831122, Guangdong Major Project of Basic and Applied Basic Research (Grant No. 2019B030302001), the Strategic Priority Research Program of the Chinese Academy of Sciences, Grant No. XDB23040100, and the Interdiscipline Research Funds of Beijing Normal University.

\section*{Data Availability}
The data underlying this article will be shared on reasonable request to the corresponding author.

\label{lastpage}
\end{document}